\documentclass[11pt]{article}
\usepackage{amsfonts}

\usepackage{graphics}
\usepackage{amsmath}
\usepackage{amssymb}
\usepackage{epsfig}
\usepackage{graphicx}
\usepackage{caption2}

\newcommand{\comparison}[1]{
\begin{center}
\begin{picture}(82,35)
\put(7,14){\framebox(7,7){$W$}} \put(28,0){\framebox(7,7){$S$}}
\put(42,0){\framebox(7,7){$K_{d}$}}
\put(56,0){\framebox(7,7){$H$}} \put(42,28){\framebox(7,7){$K$}}
\put(71,17.5){\circle{2}}

\put(0,17.5){\vector(1,0){7}} \put(14,17.5){\line(1,0){7}}
\put(72,17.5){\vector(1,0){10}}

\put(21,3.5){\vector(1,0){7}} \put(21,31.5){\vector(1,0){21}}
\put(49,31.5){\line(1,0){22}} \put(63,3.5){\line(1,0){8}}
\multiput(35,3.5)(1,0){5}{\circle*{.03}}
\multiput(49,3.5)(1,0){5}{\circle*{.03}}
\put(42,3.5){\vector(1,0){0}} \put(56,3.5){\vector(1,0){0}}
\put(21,3.5){\line(0,1){28}}
 \put(71,31.5){\vector(0,-1){13}}
\put(71,3.5){\vector(0,1){13}}  \put(5,19.5){\makebox(0,0){$u$}}
\put(80,19.5){\makebox(0,0){$z$}}
\put(73,20.5){\makebox(0,0){$-$}}
\end{picture}
\end{center}
}

\renewcommand{\captionlabeldelim}{.}
\newtheorem{example}{Example}
\newtheorem{theorem}{Theorem}
\newtheorem{corollary}{Corollary}
\newtheorem{proposition}{Proposition}

\newtheorem{definition}{Definition}
\newtheorem{remark}{Remark}

\renewcommand{\captionlabeldelim}{.}

\begin{document}

\title{Dynamical Analysis of a Networked Control System}
\author{Guofeng Zhang\thanks{School of Automation, Hangzhou Dianzi University,
Hangzhou, Zhejiang 310038, P. R. China} \and Tongwen Chen\thanks{Department of Electrical and Computer Engineering,
University of Alberta, Edmonton, Alberta, Canada T6G 2V4} \and  Guanrong Chen\thanks{Department of Electronic Engineering,
City University of Hong Kong, Hong Kong, P. R. China} \and Maria D'Amico\thanks{Depto. de Ingenier«õa El«ectrica y de Computadoras,
Universidad Nacional del Sur, Avda Alem 1253,
B8000CPB Bah«õa Blanca, Argentina} }

\maketitle

\begin{abstract}
A new network data transmission strategy was proposed in Zhang \&
Chen [2005] (arXiv:1405.2404), where the resulting nonlinear system was analyzed and
the effectiveness of the transmission strategy was demonstrated via
simulations. In this paper, we further generalize the results of
Zhang \& Chen [2005] in the
following ways: 1) Construct first-return maps of the nonlinear systems formulated in %
Zhang \& Chen [2005] and derive several existence conditions of
periodic orbits and study their properties. 2) Formulate the new
system as a hybrid system, which will ease the succeeding analysis.
3) Prove that this type of hybrid systems is not structurally stable
based on phase transition which can be applied to higher-dimensional
cases effortlessly. 4) Simulate a higher-dimensional model with
emphasis on their rich dynamics. 5) Study a class of continuous-time
hybrid systems as the counterparts of the discrete-time systems
discussed above. 6) Propose new controller design methods based on
this network data transmission strategy to improve the performance
of  each individual system and the whole network. We hope that this
research and the problems posed here will rouse interests of
researchers in such fields as control, dynamical systems and
numerical analysis.

\noindent \textbf{Keywords}: bifurcation, computational complexity,
first-return map, hybrid system, networked control system,
stability.
\end{abstract}


\newpage
\listoffigures

\newpage

\addtolength{\baselineskip}{0.4\baselineskip}

\section{Introduction}

\begin{figure}[tbh]
\epsfxsize=6in
\par
\epsfclipon
\par
\centerline{\epsffile{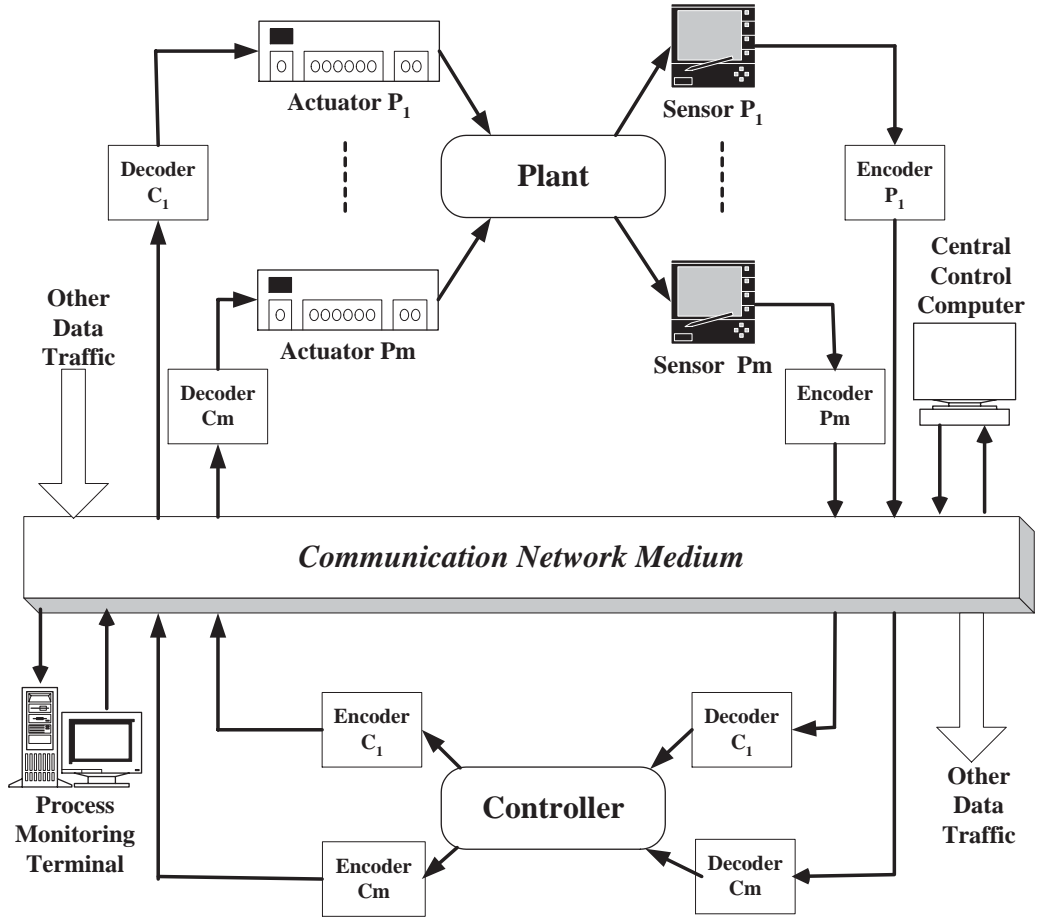}} \caption{A standard networked
control system} \label{Figure1}
\end{figure}

Consider the networked control system shown in Fig.~\ref{Figure1}.
It is obvious that connecting various system components via
communication media can reduce wiring, ease installation and improve
maintenance, among others. As far as distributed control systems are
concerned, communication among individual controllers provides each
of them with more information so that better control performance can
be achieved [Ishii \& Francis, 2002]. These advantages endorse the
network control technology a promising future in systems engineering
and applications.

Unfortunately, since the encoded system output, controller output
and other information are transmitted via communication networks
shared by many users, data traffic congestion is always unavoidable.
This usually give rise to time delays, packet loss and other
undesirable behaviors to the control systems. These problem have
become a major subject of research in this and several closely
related fields. Many network protocols and control strategies have
already been proposed to tackle the problems. Loosely speaking,
these considerations fall into three categories, which are further
discussed as follows.

The first category simply models a networked control system as a
control system with bounded time delays. In a series of recent
papers [Walsh {\it et al.}, 1999, 2001, 2002, 2002b], [Walsh \& Ye,
2001], the try-once-discard (TOD) protocol is proposed and studied
intensively, where an upper bound of sensor-to-controller time
delays induced by the network is derived, for which exponential
stability of the closed-loop system is guaranteed. This idea is
further generalized in Nesic \& Teel [2004] to derive a set of
Lyapunov UGES (Uniformly Globally Exponentially Stable) protocols in
the $L^{p}$ framework. In Yue {\it et al.} [2005], assuming bounded
time delays and packet dropouts, a robust $H_{\infty}$ control
problem is studied for networked control systems. In general this
approach is quite conservative, as has been widely acknowledged.

The second category models network time delays and packet dropouts
as random processes, typically Markov chains. In this way, some
specific features of these random processes can be utilized to
design controllers that guarantee desired system performance. In
Krtolica {\it et al.} [1994], a random model of time delays based on
Markov chain is established via augmentation. Necessary and
sufficient conditions for zero-state mean-square exponential
stability have been derived for this system. In Nilsson {\it et al.}
[1998], both sensor-to-controller and controller-to-actuator time
delays are modeled as independent white-noise with zero mean and
unit variance and consequently a (sub)optimal stochastic control
problem is studied. Two Matlab toolboxes, Jitterbug and TrueTime,
are introduced in Cervin {\it et al.} [2003], based on the principle
that networked control systems can be viewed as delayed sampled-data
systems with quantization effects. These two toolboxes can be used
as experiment platforms for research on real-time dynamical control
systems. They can be easily employed to quickly determine how
sensitive a control system is to delays, jitters, lost samples, etc.

These two categories of methods deal with network effects passively,
i.e., they solely consider the effects of network traffic on the
control systems separately, leaving aside the interactions of the
control systems and communication network. This latter consideration
is very important, which leads to the third category of
methodologies. This approach takes into account the tradeoff between
data rate and control performance. In order to minimize bandwidth
utilization, Goodwin {\it et al.} [2004] proposed some methods of
using quantization to reduce the size of the transmitted data and
solved the problem via a moving horizon technique. In Wong \&
Brockett [1999], the effect of quantization error, quantization, and
propagation time on the containability, a weaker stability concept,
of networked control systems is studied. In Takikonda \& Mitter
[2004], the tradeoff of data rate and desirable control objectives
is considered with emphasis on observability and stabilizability
under communication constraints. A necessary condition is
established on the rate for asymptotic observability and
stabilizability of a linear discrete-time system. More specifically,
the rate must be bigger than the summation of the logarithms of
modules of the unstable system poles. Then, these results are
further generalized to the study of control over noisy channels in
Takikonda \& Mitter [2004b]. The problem of asymptotic stabilization
is considered in Brockett \& Liberzon [2000], where time-varying
quantizers are designed to achieve the stabilization of an unstable
system. For the LQG optimal control of an unstable scalar system
over an additive white Gaussian noise (AWGN) channel, it is reported
[Elia, 2004] that the achievable transmission rate is given by the
Bode sensitivity integral formula, thereby establishing the
equivalence between feedback stabilization through an analog
communication channel and a communication scheme based on feedback
which unifies the design of control systems and communication
channels.

In this paper, we continue the study of the data transmission
strategy proposed in our earlier paper [Zhang \& Chen, 2005], where
a new network data transmission strategy was proposed to reduce
network traffic congestion. By
adding constant deadbands to both the controller and the plant shown in Fig.~%
\ref{Figure1}, signals will be sent only when it is necessary. By
adjusting the deadbands, a tradeoff between control performance and
reduction of network data transmission rate can be achieved. The
data transmission strategy proposed is suitable for fitting a
control network into an integrated communication network composed of
control and data networks, to fulfill the need for a new breed
geared toward total networking (see [Raji, 1994]). This problem is
of course very appealing as depicted by Raji [1994; and at the same
time it is fundamentally important so is listed in Murray {\it et
al.} [2003] as a future direction in control research in an
information-rich world: ``Current control systems are almost
universally based on synchronous, clocked systems, so they require
communication networks that guarantee delivery of sensor, actuator,
and other signals with a known, fixed delay. Although current
control systems are robust to variations that are included in the
design process (such as a variation in some aerodynamic coefficient,
motor constant, or moment of inertia), they are not at all tolerant
of (unmodeled) communication delays or dropped or lost sensor or
actuator packets. Current control system technology is based on a
simple communication architecture: all signals travel over
synchronous dedicated links, with known (or worst-case bounded)
delays and no packet loss. Small dedicated communication networks
can be configured to meet these demanding specifications for control
systems, but a very interesting question is whether we can develop a
theory and practice for control systems that operate in a
distributed, asynchronous, packet-based environment.''

Essentially speaking, under the network data transmission strategy
proposed here, in an integrated network composed of data and control
networks, it is asked that the network should provide sufficient
communication bandwidth upon request of control systems. As a
payoff, control systems will save network resources by deliberately
dropping packets without degrading system performance severely. This
is a crucial tradeoff. On the one hand, control signals are normally
time critical, hence the priority should be given to them whenever
requested; on the other hand, due to one characteristic of control
networks, namely, small packet size but frequent packet flows, it is
somewhat troublesome to manage because it demands frequent
transmissions. Our scheme aims to relieve this burden for the whole
communication network.

As we proceed, readers will find that the simple network transmission data
strategy analyzed here gives rise to many unexpected and interesting
dynamical phenomena and mathematical problems, which have innocent
appearance but are hard to deal with. More specifically, we investigate the
following issues: the stability of the control systems under the proposed
scheme; existence of periodic orbits by means of first-return maps;
bifurcation and phase transition phenomena; computational complexity. Since
this research project is oriented toward the study of control problems in a
network setting, the effectiveness of the scheme and the corresponding
controller design problem will also be addressed after system analysis.

The layout of this paper is as follows: the proposed network
protocol is presented in Sec.~2, where its advantages are discussed.
The resulting closed-loop system under this network protocol is
analyzed in Secs.~3-7. More precisely, Sec.~3 contains a study of a
closed-loop system consisting of a scalar plant controlled by a
proportional controller with a constant gain, as the simplest case
under this framework to provide a detailed analysis. The structural
stability of the system is studied in Sec.~4. Sec.~5 studies the
rich dynamics of a higher-dimensional model. The continuous-time
counterpart of this type of discrete-time systems is discussed in
Sec.~5. The controller design problem is then addressed in Sec.~7.
Some concluding remarks, open problems and future research issues
are finally posed and discussed in Sec.~8.

\section{The Proposed Network Protocol}

In Zhang \& Chen [2005], a new data transmission strategy was
proposed, which is briefly reviewed here. Consider the feedback system shown in Fig.~\ref%
{Figure2},
\begin{figure}[tbh]
\epsfxsize=4in
\par
\epsfclipon
\par
\centerline{\epsffile{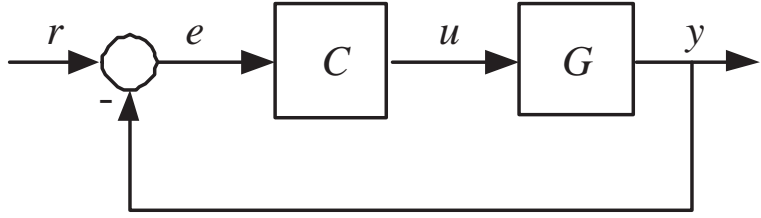}} \caption{A typical feedback
system} \label{Figure2}
\end{figure}
where $G$ is a discrete-time system of the form:
\begin{eqnarray}
x(k+1) &=&Ax(k)+Bu(k),  \label{sysG} \\
y(k) &=&Cx(k),  \notag
\end{eqnarray}
with the state $x\in \mathbb{R}^{n}$, the input $u\in \mathbb{R}^{m}$, the
output $y\in \mathbb{R}^{p}$, and the reference input $r\in \mathbb{R}^{p}$,
respectively; $C$ is a stabilizing controller:
\begin{eqnarray}
x_{d}(k+1) &=&A_{d}x_{d}(k)+B_{d}e(k),  \label{conC} \\
u(k) &=&C_{d}x_{d}(k)+D_{d}e(k),  \notag \\
e\left( k\right) &=&r\left( k\right) -y\left( k\right) ,  \notag
\end{eqnarray}
with its state $x_{d}\in \mathbb{R}^{n_{c}}$. Let $\xi =\left[
\begin{array}{c}
x \\
x_{d}%
\end{array}
\right] $. Then, the closed-loop system from $r$ to $e$ is described by
\begin{eqnarray}
\xi \left( k+1\right) &=&\left[
\begin{array}{cc}
A-BD_{d}C & BC_{d} \\
-B_{d}C & A_{d}%
\end{array}
\right] \xi \left( k\right) +\left[
\begin{array}{c}
BD_{d} \\
B_{d}%
\end{array}
\right] r(k),  \label{clsys1} \\
e(k) &=&\left[
\begin{array}{cc}
-C & 0%
\end{array}
\right] \xi \left( k\right) +r(k).  \notag
\end{eqnarray}

\begin{figure}[tbh]
\epsfxsize=4in
\par
\epsfclipon
\par
\centerline{\epsffile{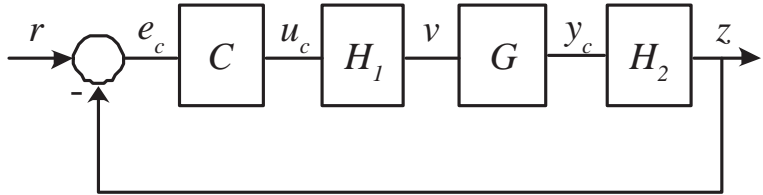}} \caption{A constrained feedback
system} \label{Figure3}
\end{figure}

Now, we add some nonlinear constraints on both $u$ and $y$.
Specifically, consider the system shown in Fig.~\ref{Figure3}.
The nonlinear constraint $H_{1}$ is defined as follows: for a given $\delta
_{1}>0$, let $v(-1)=0$; and for $k\geq0$, let
\begin{equation}
v(k)=H_{1}\left( u_{c}\left( k\right) ,v(k-1)\right) =\left\{
\begin{array}{ll}
u_{c}(k), & \mbox{if~}\left\| u_{c}\left( k\right) -v\left( k-1\right)
\right\| _{\infty }>\delta _{1}, \\
v(k-1), & \mbox{otherwise.}%
\end{array}
\right.  \label{constraint1}
\end{equation}
Similarly, $H_{2}$ is defined as follows: for a given $\delta _{2}>0$, let $%
z(-1)=0$; for $k\geq0$, let
\begin{equation}
z(k)=H_{2}\left( y_{c}\left( k\right) ,z(k-1)\right) =\left\{
\begin{array}{ll}
y_{c}(k), & \mbox{if~}\left\| y_{c}\left( k\right) -z\left( k-1\right)
\right\| _{\infty }>\delta _{2}, \\
z(k-1), & \mbox{otherwise.}%
\end{array}
\right.  \label{constraint2}
\end{equation}
It can be shown that $\left\| H_{1}\right\| $, the induced norm of $H_{1}$,
equals $2$, and so is $\left\| H_{2}\right\| $.

In Octanez {\it et al.} [2002] \textit{adjustable} deadbands are
proposed to reduce network traffics, where the closed-loop system
with deadbands is modeled as a perturbed system, with exponential
stability followed from that of the original system [Khalil, 1996].
The constraints, $\delta _{1}$ and $\delta _{2}$, proposed here are
fixed. We have observed [Zhang \& Chen, 2005] that the stability of
the system shown in Fig.~\ref{Figure3} is fairly complicated and
only local stability can be obtained. However, the main advantage of
fixed deadbands is that it will reduce network traffics more
effectively. Furthermore, the stability region can be scaled as
large as desired.

For the ``constrained'' system shown in Fig.~\ref{Figure2}, let $p$
denote the state of the system $G$ and $p_{d}$ denote the state of
the controller $C$. Then
\begin{eqnarray*}
p(k+1) &=&Ap(k)+Bv(k), \\
y_{c}(k) &=&Cp(k),
\end{eqnarray*}
and
\begin{eqnarray*}
p_{d}(k+1) &=&A_{d}p_{d}(k)+B_{d}e_{c}(k), \\
u_{c}(k) &=&C_{d}p_{d}(k)+D_{d}e_{c}(k), \\
e_{c}(k) &=&r(k)-z(k).
\end{eqnarray*}
Let $\eta =\left[
\begin{array}{c}
p \\
p_{d}%
\end{array}
\right] $. Then, the closed-loop system from $r$ to $e$ is
\begin{eqnarray}
\eta (k+1) &=&\left[
\begin{array}{cc}
A & 0 \\
0 & A_{d}%
\end{array}
\right] \eta (k)+\left[
\begin{array}{cc}
B & 0 \\
0 & B_{d}%
\end{array}
\right] \left[
\begin{array}{c}
v(k) \\
-z(k)%
\end{array}
\right] +\left[
\begin{array}{c}
0 \\
B_{d}%
\end{array}
\right] r(k),  \label{clsys2} \\
e_{c}(k) &=&\left[
\begin{array}{cc}
-C & 0%
\end{array}
\right] \eta \left( k\right) +r(k),  \notag
\end{eqnarray}
where $v$ and $z$ are given in Eqs. (\ref{constraint1})-(\ref{constraint2}).

At this point, one can see that in the framework of the
communication network containing both data and control networks,
this proposed data transmission strategy will provide sufficient
communication bandwidth upon request of control networks used by the
control systems. As a payoff, the control systems will save some
network resources by deliberately dropping packets. This
consideration is well tailored to the requirement of control
networks in general. On the one hand, control signals are normally
time critical, hence the priority should be given to them whenever
requested. On the other hand, due to the characteristics of control
networks, namely, small packet size but frequent packets flows, it
is somewhat troublesome to manage because it demands frequent
transmissions. Our scheme aims to relieve this burden for the whole
communication network.

\section{First-Return Maps}

To simplify the following discussions, suppose that the system $G$ in Fig.~%
\ref{Figure2} is a scalar system, the controller $C$ is simply -1,
there is no $H_{2}$ involved, and $r=0$. That is, in this section,
we consider the following simplified system:
\begin{equation}
x(k+1)=ax(k)+bv(k),  \label{original system}
\end{equation}%
with $v(-1)\in \mathbb{R}$, and for $k\geq 0$,
\begin{eqnarray}
v\left( k\right) &=&H\left( x\left( k\right) ,v\left( k-1\right) \right)
\notag \\
&:=&\left\{
\begin{array}{ll}
x\left( k\right), & \mbox{if~}\left| x\left( k\right) -v\left( k-1\right)
\right| >\delta , \\
v\left( k-1\right), & \mbox{otherwise,}%
\end{array}%
\right.  \label{original_constraint}
\end{eqnarray}%
where $|a+b|<1$ and $\delta $ is a positive number.

In Zhang \& Chen [2005], the system composed of Eqs. (\ref{original system})-(\ref%
{original_constraint}) was studied in great detail, where a
necessary condition for the existence of periodic orbits was
derived. We now generalize it and provide a necessary and sufficient
condition and give a characterization of the state space based on
it. This analysis is important: To achieve good control, a nonlinear
system may be desired to work near an equilibrium point or a limit
cycle. In the case that a limit cycle is preferred, this result will
reveal under what condition limit cycles may exist and starting from
where a trajectory may converge to the desired limit cycle. In the
case that an equilibrium is desirable, this result will provide the
designer with some information as how to design controllers to
prevent trajectories from being stuck into a limit cycle. Hence,
this analysis will provide useful insights into the design of
control systems under the proposed data transmission strategy. We
now investigate this important problem case by case.

\subsection{Case 1: $0<a<1$, $b<0$}

\begin{figure}[tbp]
\epsfxsize=4in
\par
\epsfclipon
\par
\centerline{\epsffile{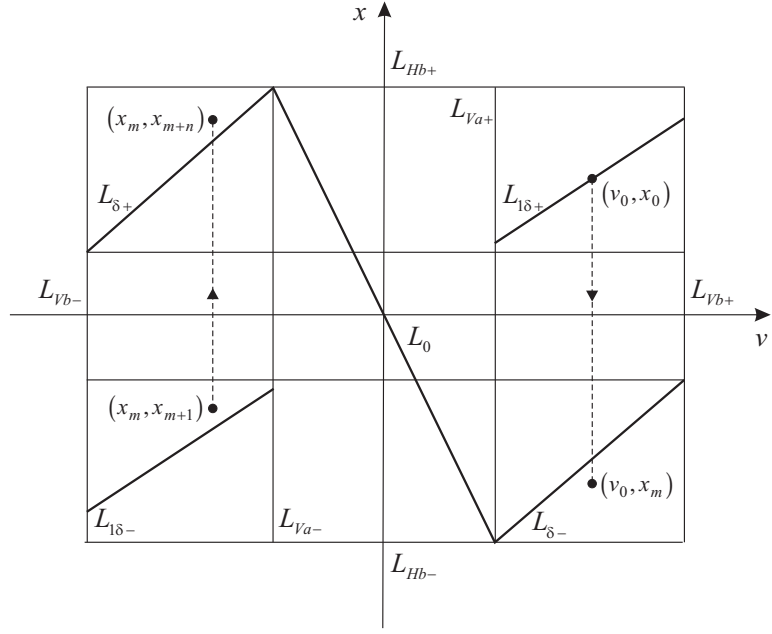}} \caption{Diagram for the case
with $0<a<1$ and $b<0$} \label{Figure4}
\end{figure}

For convenience, we present Fig.~\ref{Figure4},
where
\begin{eqnarray*}
L_{o} &:=&\left\{ \left( v_{-},x\right) \in I_{a}:x=\frac{b}{1-a}%
v_{-}\right\} , \\
L_{\delta +} &:=&\left\{ \left( v_{-},x\right) \in I_{b}:x-v_{-}=\delta
\right\} , \\
L_{\delta -} &:=&\left\{ \left( v_{-},x\right) \in I_{b}:x-v_{-}=-\delta
\right\} \\
L_{1\delta +} &:=&\left\{ \left( v_{-},x\right) \in I_{b}:x=\left(
a+b\right) v_{-} ,v_{-}>\frac{1-\left| a\right| }{1-\left( a+b\right) }%
\delta \right\} , \\
L_{1\delta -} &:=&\left\{ \left( v_{-},x\right) \in I_{b}:x=\left(
a+b\right) v_{-} ,v_{-}<-\frac{1-\left| a\right| }{1-\left( a+b\right) }%
\delta \right\}, \\
L_{Hb+} &:=&\left\{ \left( v_{-},\frac{-b}{1-\left| a+b\right| }\delta
\right) :\left| v_{-}\right| \leq \frac{-b}{1-\left| a+b\right| }\delta
\right\} , \\
L_{Hb-} &:=&\left\{ \left( v_{-},\frac{b}{1-\left| a+b\right| }\delta
\right) :\left| v_{-}\right| \leq \frac{-b}{1-\left| a+b\right| }\delta
\right\} , \\
L_{Vb+} &:=&\left\{ \left( \frac{-b}{1-\left| a+b\right| }\delta ,x\right)
:\left| x\right| \leq \frac{-b}{1-\left| a+b\right| }\delta \right\} , \\
L_{Vb-} &:=&\left\{ \left( \frac{b}{1-\left| a+b\right| }\delta ,x\right)
:\left| x\right| \leq \frac{-b}{1-\left| a+b\right| }\delta \right\}, \\
L_{Va+} &:=&\left\{ \left( \frac{1-\left| a\right| }{1-\left( a+b\right) }%
\delta ,x\right) ,:\left| x\right| \leq \frac{-b}{1-\left( a+b\right) }%
\delta \right\} , \\
L_{Va-} &:=&\left\{ \left( -\frac{1-\left| a\right| }{1-\left( a+b\right) }%
\delta ,x\right) ,:\left| x\right| \leq \frac{-b}{1-\left( a+b\right) }%
\delta \right\}.
\end{eqnarray*}%
Assume that an initial condition $(v_{0},x_{0})$ is located on the line $%
L_{1\delta +}$ satisfying
\begin{equation*}
x_{0}=(a+b)v_{0}.
\end{equation*}
Moreover, suppose that the trajectory starting from it does not
converge to a fixed point (see [Zhang \& Chen, 2005] for details).
Hence, the successive iterations are given by
\begin{eqnarray*}
x_{1} &=&ax_{0}+bv_{0}=\left( a^{2}+ab+b\right) v_{0}=\left(
a^{2}+\sum_{i=0}^{1}a^{i}b\right) v_{0}, \\
v_{1} &=&v_{0},
\end{eqnarray*}%
\begin{eqnarray*}
x_{2} &=&ax_{1}+bv_{0}=\left( a^{3}+\sum_{i=0}^{2}a^{i}b\right) v_{0}, \\
v_{2} &=&v_{0},
\end{eqnarray*}
\begin{equation*}
\vdots
\end{equation*}%
\begin{eqnarray*}
x_{m} &=&ax_{m-1}+bv_{0}=\left( a^{m+1}+\sum_{i=0}^{m}a^{i}b\right) v_{0}, \\
v_{m} &=&v_{0}.
\end{eqnarray*}%
Accordingly,%
\begin{equation*}
v_{m}-x_{m}=\left( 1-a^{m+1}-\sum_{i=0}^{m}a^{i}b\right) v_{0}.
\end{equation*}
As indicated in Fig.~\ref{Figure4}, the orbit moves downward
following the line $v = v_0$ on the right part of the region. In
this way, there will
exist a value of $m$ such that the trajectory crosses the line segment $%
L_{\delta-}$, i.e.,
\begin{equation}
\left| x_{m}-v_{m}\right| >\delta ,  \label{chaos-1-switch}
\end{equation}%
and note that such an $m$ always exists. Thus%
\begin{eqnarray*}
\left( 1-a^{m+1}-\sum_{i=0}^{m}a^{i}b\right) v_{0} >\delta & \Leftrightarrow
& \frac{\left( 1-a\right) \delta }{(1-(a+b))v_{0}}<1-a^{m+1} \\
&\Leftrightarrow &a^{m+1}<1-\frac{\left( 1-a\right) \delta }{(1-(a+b))v_{0}}
\\
&\Leftrightarrow &m>\frac{\ln \left( 1-\frac{\left( 1-a\right) \delta }{%
(1-(a+b))v_{0}}\right) }{\ln a}-1.
\end{eqnarray*}%
Hence, the smallest $m$ is given by
\begin{equation}
m=\left\lceil \frac{\ln \left( 1-\frac{\left( 1-a\right) \delta }{%
(1-(a+b))v_{0}}\right) }{\ln a}\right\rceil -1,  \label{m}
\end{equation}%
where $\lceil r \rceil$ is the least integer bigger than $r$. Note that
\begin{eqnarray*}
x_{m+1} &=&\left( a+b\right) x_{m}, \\
v_{m+1} &=&x_{m},
\end{eqnarray*}
Since $x_{m}<0$, this point is located on the left part of Fig.~\ref%
{Figure4}, and also
\begin{equation*}
\left| x_{m+1}-v_{m+1}\right| <\delta .
\end{equation*}
Hence
\begin{eqnarray*}
x_{m+2} &=&ax_{m+1}+bv_{m}=\left( a^{2}+\sum_{i=0}^{1}a^{i}b\right) x_{m}, \\
v_{m+2} &=&x_{m},
\end{eqnarray*}%
\begin{equation*}
\vdots
\end{equation*}%
\begin{eqnarray*}
x_{m+n} &=&\left( a^{n}+\sum_{i=0}^{n-1}a^{i}b\right) x_{m}, \\
v_{m+n} &=&x_{m},
\end{eqnarray*}%
\begin{equation*}
x_{m+n}-v_{m+n}=\left( a^{n}+\sum_{i=0}^{n-1}a^{i}b-1\right) x_{m}.
\end{equation*}%
The orbit now moves upward along the line $v=x_{m}$ (see Fig.~\ref{Figure4}%
). Then, there will exist a value of $n$ such that the trajectory crosses $%
L_{\delta +}$, i.e.,
\begin{equation*}
\left| x_{m+n}-v_{m+n}\right| >\delta.
\end{equation*}%
Note that%
\begin{equation*}
v_{m+n}=x_{m}<0.
\end{equation*}%
Then%
\begin{eqnarray*}
\left( a^{n}+\sum_{i=0}^{n-1}a^{i}b-1\right) x_{m} &=&\left(
a^{n}+\sum_{i=0}^{n-1}a^{i}b-1\right) \left(
a^{m+1}+\sum_{i=0}^{m}a^{i}b\right) v_{0}>\delta \\
&\Leftrightarrow &1-a^{n}>\frac{\left( 1-a\right) \delta }{(a+b-1)\left(
\left( 1-a^{m+1}\right) \frac{a+b-1}{1-a}+1\right) v_{0}} \\
&\Leftrightarrow &n>\frac{\ln \left( 1-\frac{\left( 1-a\right) \delta }{%
(a+b-1)\left( \left( 1-a^{m+1}\right) \frac{a+b-1}{1-a}+1\right) v_{0}}%
\right) }{\ln a}.
\end{eqnarray*}%
Hence, the smallest $n$ is
\begin{equation}
n=\left\lceil \frac{\ln \left( 1-\frac{\left( 1-a\right) \delta }{%
(a+b-1)\left( \left( 1-a^{m+1}\right) \frac{a+b-1}{1-a}+1\right) v_{0}}%
\right) }{\ln a}\right\rceil .  \label{n}
\end{equation}%
The switching law (Eq. \ref{original_constraint}) provokes that the new
iteration point
\begin{eqnarray*}
x_{m+n+1} &=&\left( a+b\right) x_{m+n}, \\
v_{m+n+1} &=&x_{m+n},
\end{eqnarray*}%
returns to the zone where the trajectory was originated. In particular, if
\begin{equation*}
\left( v_{m+n+1},x_{m+n+1}\right) =\left( v_{0},x_{0}\right)
\end{equation*}
then one will get a closed orbit. This motivates us to define the first-return map%
\begin{eqnarray}
\varphi &:&L_{1}\rightarrow L_{1}  \notag \\
v &\mapsto &\left( a^{n}+\sum_{i=0}^{n-1}a^{i}b\right) \left(
a^{m+1}+\sum_{i=0}^{m}a^{i}b\right) v,  \label{poincare_map}
\end{eqnarray}%
where $L_{1}$ is the projection of $L_{1\delta +}$ onto the $v$
axis, and $m$ and $n$ satisfy Eqs. (\ref{m})-(\ref{n}),
respectively.

\begin{definition}
\label{chaos-1-firstreturn-def} (Type 1 periodic orbits) A periodic orbit
starting from $\left( v_{0},x_{0}\right)\in L_{1\delta +}$ is said to be of
type 1 if
\begin{equation}
\varphi \left( v_{0}\right) =v_{0},  \label{criterion_periodic}
\end{equation}%
where $\varphi $ is defined by Eq. (\ref{poincare_map}), and $m$ and
$n$ satisfy Eqs. (\ref{m})-(\ref{n}), respectively. In this case,
the period of this orbit starting from $\left( v_{0},x_{0}\right) $
is $m+n+1$.
\end{definition}

\begin{remark}
\label{chaos-1-firstreturn-rema-type1} {\rm A periodic orbit is of
type 1 if it forms a closed loop right after the first return. There
are possibly other periodic orbits that become closed loops after
several returns. These orbits can be studied in a similar way, but
it is more computationally involved.}
\end{remark}

The following result follows immediately from the foregoing discussions.

\begin{theorem}
\label{chaos-1-firstreturn-thm-type1} The trajectory starting from $\left(
v_0,x_0\right) $ is periodic of type 1 if and only if Eq. (\ref%
{criterion_periodic}) holds.
\end{theorem}

Actually, we can find all periodic orbits of type 1: If Eq. (\ref%
{criterion_periodic}) holds, i.e.,
\begin{equation*}
\varphi \left( v_0\right) =v_0,
\end{equation*}%
then%
\begin{equation*}
\left( a^{n}+\sum_{i=0}^{n-1}a^{i}b\right) \left(
a^{m+1}+\sum_{i=0}^{m}a^{i}b\right) =1,
\end{equation*}%
i.e.,
\begin{equation}
\left( \left( 1-a^{n}\right) \frac{a+b-1}{1-a}+1\right) \left( \left(
1-a^{m+1}\right) \frac{a+b-1}{1-a}+1\right) =1  \label{periodic_orbits}
\end{equation}%
for some $m,n>0$. Given that $0<a<1$, $b<0$ and $\left| a+b\right|
<1$, $m$ and $n$ satisfying (\ref{periodic_orbits}) are both finite.
Hence, all periodic orbits of type 1 can be found.

\begin{remark}
\label{chaos-1-firstreturn-dense-periodic} {\rm If $\left(
v,x\right) $
leads to a periodic orbit of type 1, according to Eqs (\ref{m}) and (\ref{n}%
), there exists a neighborhood of $\left( v,x\right) $ on
$L_{1\delta +}$ such that each point of which will lead to a
periodic orbit of type 1, so all such orbits are together dense.}
\end{remark}

\subsection{Case 2: $a=1$}

Assume that an initial condition $\left( v,x\right) $ satisfies%
\begin{equation*}
x=\left( 1+b\right) v, v>0,
\end{equation*}%
and also suppose that the orbit starting from it is within the oscillating
region. Then
\begin{eqnarray*}
x_{1} &=&x+bv=\left( 1+2b\right) v, \\
v_{1} &=&v,
\end{eqnarray*}%
\begin{equation*}
\vdots
\end{equation*}
\begin{eqnarray*}
x_{m} &=&x_{1}+bv_{1}=\left( 1+\left( m+1\right) b\right) v, \\
v_{m} &=&v.
\end{eqnarray*}%
Suppose that%
\begin{equation*}
v_{m}-x_{m}=-\left( m+1\right) bv>\delta .
\end{equation*}%
Then%
\begin{equation*}
m+1>\frac{\delta }{\left( -b\right) v}.
\end{equation*}%
Hence, the least $m$ is given by
\begin{equation*}
m=\left\lceil \frac{\delta }{\left( -b\right) v}\right\rceil -1.
\end{equation*}%
Moreover,
\begin{eqnarray*}
x_{m+1} &=&\left( 1+b\right) x_{m}, \\
v_{m+1} &=&x_{m}<0,
\end{eqnarray*}
\begin{equation*}
\vdots
\end{equation*}
\begin{eqnarray*}
x_{m+n} &=&\left( 1+nb\right) x_{m}, \\
v_{m+n} &=&x_{m}.
\end{eqnarray*}%
Suppose that
\begin{equation*}
x_{n+m}-v_{n+m}=nbx_{m}>\delta .
\end{equation*}%
Then%
\begin{equation*}
n>\frac{\delta }{bx_{m}}=\frac{\delta }{b\left( 1+\left( m+1\right) b\right)
v}.
\end{equation*}%
Hence, the smallest $n$ is given by%
\begin{equation*}
n=\left\lceil \frac{\delta }{b\left( 1+\left( m+1\right) b\right) v}%
\right\rceil .
\end{equation*}%
Define%
\begin{eqnarray}
\varphi &:&L_{1}\rightarrow L_{1}  \notag \\
v &\mapsto &\left( 1+nb\right) \left( 1+\left( m+1\right) b\right) v,
\label{Poinc_map_2}
\end{eqnarray}
If%
\begin{equation*}
\varphi \left( v\right) =v,
\end{equation*}%
then%
\begin{equation}
\frac{1}{m+1}+\frac{1}{n}=-b.  \label{peirodic}
\end{equation}

\begin{theorem}
\label{chaos-1-firstreturn-a=1-thm} The trajectory starting from $\left(
v,(1+b)v\right) $ is periodic of type 1 if and only if $v$ is a fixed point
of the first-return map defined in Eq. (\ref{Poinc_map_2}).
\end{theorem}

\begin{remark}
\label{chaos-1-firstreturn-a=1-rem} {\rm This result is a
generalization of Theorem 3 in Zhang \& Chen [2005], where the
condition is only necessary. For example, given $a=1$ and $b=-1/2$,
the origin is the unique invariant set.
It is obvious that $p=q=4$ is a solution to%
\begin{equation*}
\frac{1}{p}+\frac{1}{q}=-b.
\end{equation*}%
However, there are no periodic orbits. This indicates that the
necessary condition given by Theorem 3 in Zhang \& Chen [2005] is
not sufficient.}
\end{remark}

Next we find all periodic orbits of type 1 for the case of $a=1$.

For convenience, here we use $m$ instead of $m+1$ in Eq. (\ref{peirodic}).
Suppose%
\begin{equation*}
b=-\frac{q}{p},
\end{equation*}%
where $p>0$, $q>0$, $\gcd \left( p,q\right) =1$. According to Eq. (\ref%
{peirodic}),
\begin{equation*}
\frac{1}{m}=-b-\frac{1}{n}=\frac{pn-q}{qn},
\end{equation*}%
i.e.,
\begin{equation*}
m=\frac{qn}{pn-q}.
\end{equation*}%
Obviously,%
\begin{equation*}
m>\frac{q}{p}.
\end{equation*}%
Furthermore, $m$ is a decreasing function of $n$. By symmetry, let
\begin{equation*}
n_{0}=\left\lceil \frac{q}{p}\right\rceil .
\end{equation*}%
Then%
\begin{equation*}
\left\lceil \frac{q}{p}\right\rceil \leq m\leq \left\lceil \frac{qn_{0}}{pn_{0}-q}%
\right\rceil .
\end{equation*}%
Similarly,
\begin{equation*}
\left\lceil \frac{q}{p}\right\rceil \leq n\leq \left\lceil \frac{qm_{0}}{%
pm_{0}-q}\right\rceil ,
\end{equation*}%
where%
\begin{equation*}
m_{0}=\left\lceil \frac{q}{p}\right\rceil .
\end{equation*}%
Based on this analysis and Theorem \ref{chaos-1-firstreturn-a=1-thm}, all
periodic orbits of type 1 can be determined.

\subsection{Case 3 $a=-1$ and $\left| a+b\right| <1$}

For this case, each trajectory is an eventually periodic orbit of period 2.

\subsection{Case 4 $a>1$}

This is similar to the case of $0<a<1$. The only difference is%
\begin{equation*}
m<\frac{\ln \left( 1-\frac{\left( 1-a\right) \delta }{(1-(a+b))v}\right) }{%
\ln a}-1,
\end{equation*}%
due to $\ln a>0$. The least $m$ is
\begin{equation*}
m=\left\lceil \frac{\ln \left( 1-\frac{\left( 1-a\right) \delta }{(1-(a+b))v}%
\right) }{\ln a}\right\rceil -1.
\end{equation*}

The complex dynamics exhibited in this system is due to its
nonlinearity induced by switching. This is different from that of a
quantized system. The complicated behavior of an unstable quantized
scalar system has been extensively studied in Delchamps [1988, 1989,
1990] and Fagnani \& Zampieri [2003], and the MIMO case is addressed
in Fagnani \& Zampieri [2004]. In Delchamps [1990], it is mentioned
that if the system parameter $a$ is stable, a quantized system may
have many fixed points as well as periodic orbits, which are all
asymptotically stable. However, for the constrained systems here,
almost all trajectories are not periodic orbits. For systems with
$a=1$, periodic orbits are locally stable, which is not the case for
a quantized system [Delchamps, 1990]. Given that $a$ is unstable,
the ergodicity of the quantized system is investigated in Delchamps
[1990]. In essence, related results there depend heavily on the
affine representation of the system by which the system is piecewise
expanding, i.e., the absolute value of the derivative of the
piecewise affine map in each interval partitioned naturally by it is
greater than $1$. Based on this crucial property, the main theorem
(Theorem 1) in Lasota \& Yorke [1973] and then that in Li \& Yorke
[1978] are employed to show that there exists a unique invariant
measure under the affine map on which the map is also ergodic.
Therefore, ergodicity has been established for scalar unstable
quantized systems. However, this is not the case for the system
studied here. Though the system is still piecewise linear, it is
\textbf{singular} with respect to the Lebesgue measure and,
furthermore, the derivative of the system map in a certain
region is $\left( a+b\right) $, whose absolute value is strictly less than $%
1 $. Hence, the results in Lasota \& Yorke [1973] and Li \& Yorke
[1978] are not applicable here. However, by extensive experiments,
we strongly believe that the system indeed has the property of
ergodicity. This will be left as a conjecture for future research to
verify theoretically.

\section{Structural Stability}

Loosely speaking, a nonlinear system is \textit{structurally stable} if a
slight perturbation of its system parameters will not change its phase
portrait qualitatively. In Zhang \& Chen [2005], we proved that given $a=1$ in system  (\ref%
{original system}), if there are periodic orbits, then the system is not
structurally stable. In this section, we will show that the condition of $%
a=1 $ is actually unnecessary.
\begin{figure}[tbh]
\epsfxsize=3.5in
\par
\epsfclipon
\par
\centerline{\epsffile{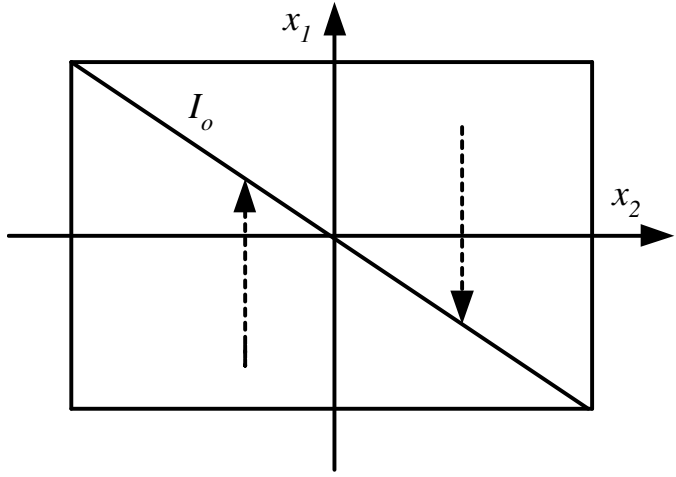}} \caption{A global attracting
region of a type-1 generic system} \label{Figure5}
\end{figure}
\begin{figure}[tbh]
\epsfxsize=3.5in
\par
\epsfclipon
\par
\centerline{\epsffile{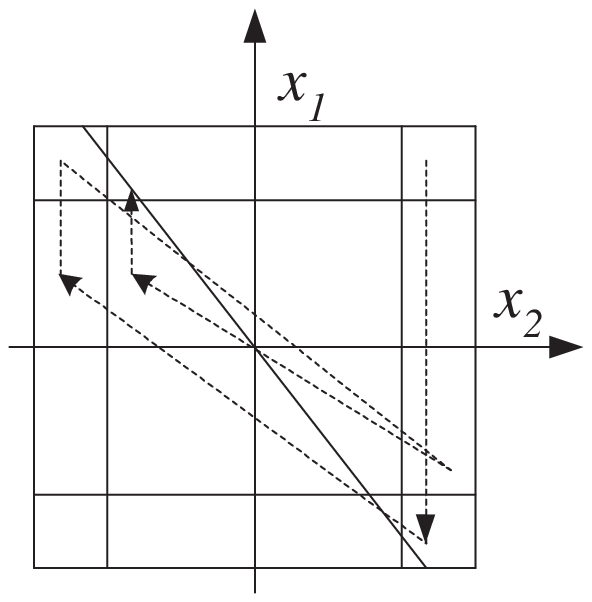}} \caption{A global attracting
region of the generic system with $a=0.3$ and $b=-0.9$}
\label{Figure6}
\end{figure}

We begin with the simplest case, namely, a generic system, whose
unique attractor is the line segment of fixed points. Even such a
simple case can still be classified into at least  two categories.
We proved in Zhang \& Chen [2005] that if the  parameters of the
system defined by Eqs. (\ref{original
system})-(\ref{original_constraint}) satisfy
\begin{equation}
\frac{1-|a|}{1-(a+b)}>\frac{|b|}{1-|a+b|} \label{unique condition}
\end{equation}%
with $|a+b|<1$ and $|a|<1$, then it is  generic. One of its global
attracting regions is shown in Fig.~\ref{Figure5}, where $I_{o}$ is
the line segment of fixed points, which runs from the point
$(x_{2},x_{1})=\left(-\frac{(1-|a|)\delta}{1-(a+b)},
\frac{b}{1-a}\frac{(1-|a|)\delta}{1-(a+b)}\right)$ on the left to
the point $(x_{2},x_{1})=\left(\frac{(1-|a|)\delta}{1-(a+b)},
-\frac{b}{1-a}\frac{(1-|a|)\delta}{1-(a+b)}\right)$ on the right. A
trajectory will converge {\it vertically} to a certain point on
$I_o$. On the other hand, we proved that the system with parameters
$a=\frac{3}{10}$ and $b=-\frac{9}{10}$, which do not satisfy Eq.
(\ref{unique condition}), is also generic whose typical trajectories
are like that shown in Fig.~\ref{Figure6} (the trajectory starting
from $(0.005,0.005)$ around converges to a fixed point close to
$(0.0004,0.004)$ after several oscillations). Till now, we have not
found a third type of generic systems. Apparently the first generic
system is simpler than the second one. Hence, we first investigate
the structural stability of the first type of systems. For
convenience, we call such systems type-1 generic systems or generic
systems of type 1. Observing that each type-1 generic system has a
global attracting region as shown in Fig.~\ref{Figure5}, thereby we
focus on its behavior in this region.

The following result asserts that two generic systems of type 1 are
`identical' in the sense of topology:

\begin{proposition}\label{paper-prop1}
 Two type-1 generic systems are homeomorphic, i.e., there exists
a bijective map form one to the other which has a continuous
inverse.
\end{proposition}

\noindent \textbf{Proof.}~~  For convenience, define%
\begin{equation*}
x_{2}\left( k\right) :=v\left( k-1\right) ,~~k\geq 0.
\end{equation*}%
Then, the original system defined in Eqs. (\ref{original system})-(\ref%
{original_constraint}) is equivalent to
\begin{equation}
\left[
\begin{array}{c}
x_{1}\left( k+1\right)  \\
x_{2}\left( k+1\right)
\end{array}%
\right] =\left\{
\begin{array}{ll}
\left[
\begin{array}{cc}
a+b & 0 \\
1 & 0%
\end{array}%
\right] \left[
\begin{array}{c}
x_{1}\left( k\right)  \\
x_{2}\left( k\right)
\end{array}%
\right] , & \mbox{if~}\left| x_{1}\left( k\right) -x_{2}\left( k\right)
\right| >\delta , \\
\left[
\begin{array}{ll}
a & b \\
0 & 1%
\end{array}%
\right] \left[
\begin{array}{c}
x_{1}\left( k\right)  \\
x_{2}\left( k\right)
\end{array}%
\right] , & \mbox{otherwise.~~~}%
\end{array}%
\right.   \label{modified system}
\end{equation}
\begin{figure}[tbh]
\epsfxsize=4in
\par
\epsfclipon
\par
\centerline{\epsffile{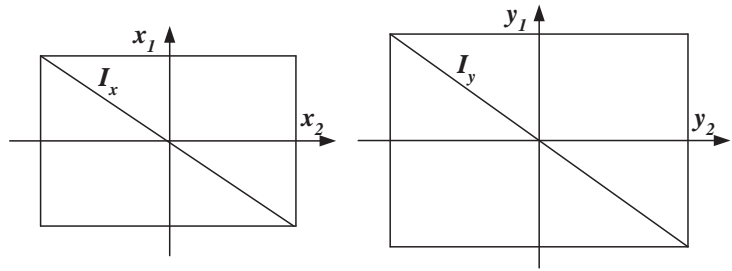}} \caption{Global attracting
regions of systems $\Sigma _{1}$ and $\Sigma _{2}$ } \label{Figure7}
\end{figure}
Consider the following two type-1 generic systems whose global
attracting regions are shown in Fig.~\ref{Figure7}:
\begin{equation*}
\Sigma _{1}:~~ x\left( k+1\right) =\left\{
\begin{array}{ll}
B_{1}x\left( k\right) , & \mbox{if~}\left| x_{1}\left( k\right) -x_{2}\left(
k\right) \right| >\delta , \\
A_{1}x\left( k\right) , & \mbox{otherwise,~~~}%
\end{array}%
\right.
\end{equation*}%
\begin{equation*}
\Sigma _{2}:~~ y\left( k+1\right) =\left\{
\begin{array}{ll}
B_{2}x\left( k\right) , & \mbox{if~}\left| y_{1}\left( k\right) -y_{2}\left(
k\right) \right| >\delta , \\
A_{2}x\left( k\right) , & \mbox{otherwise,~~~}%
\end{array}%
\right.
\end{equation*}%
in which
\begin{equation*}
x=\left[
\begin{array}{c}
x_{1} \\
x_{2}%
\end{array}%
\right] , ~ y=\left[
\begin{array}{c}
y_{1} \\
y_{2}%
\end{array}%
\right] , ~ A_{i}=\left[
\begin{array}{cc}
a_{i} & b_{i} \\
0 & 1%
\end{array}%
\right] ,~B_{i}=\left[
\begin{array}{cc}
a_{i}+b_{i} & 0 \\
1 & 0%
\end{array}%
\right] ,~~i=1,2.
\end{equation*}%
Next, we define a map $h$ from $I_{x}$ to $I_{y}$ by
\begin{eqnarray}
h &:&I_{x}\longrightarrow I_{y},  \label{projection-fixed points} \\
&&\left( \frac{b_{1}}{1-a_{1}}x_{2},x_{2}\right)  \mapsto \left( \frac{b_{2}%
}{1-a_{2}}\frac{\frac{1-\left| a_{2}\right| }{1-(a_{2}+b_{2})}}{\frac{%
1-\left| a_{1}\right| }{1-(a_{1}+b_{1})}}x_{2},\frac{\frac{1-\left|
a_{2}\right| }{1-(a_{2}+b_{2})}}{\frac{1-\left| a_{1}\right| }{%
1-(a_{1}+b_{1})}}x_{2}\right) .  \nonumber
\end{eqnarray}%
Clearly, $h$ is one-to-one, onto and has a continuous inverse.

Next, define
\begin{eqnarray}
\tilde{h} &:&\Sigma _{1}\longrightarrow \Sigma _{2},  \label{new map} \\
&&\left( x_{1},x_{2}\right)  \mapsto \left( \frac{\frac{b_{2}}{1-a_{2}}}{%
\frac{b_{1}}{1-a_{1}}}\frac{\frac{1-\left| a_{2}\right| }{1-(a_{2}+b_{2})}}{%
\frac{1-\left| a_{1}\right| }{1-(a_{1}+b_{1})}}x_{1},\frac{\frac{1-\left|
a_{2}\right| }{1-(a_{2}+b_{2})}}{\frac{1-\left| a_{1}\right| }{%
1-(a_{1}+b_{1})}}x_{2}\right) .  \nonumber
\end{eqnarray}%
It is easy to see that the projection of $\tilde{h}$ on $I_{x}$ is
exactly $h $, and furthermore $\tilde{h}$ is a homeomorphism.
Actually system $\Sigma _{2}$ can be obtained by stretching (or
contracting) system $\Sigma _{1}$, therefore they are topologically
equivalent. $\hfill$ $\blacksquare$

Two remarks are in order.

\begin{remark}{\rm
It is hard to apply the foregoing method to other types of generic
systems because they may not have such simple global attracting
regions.}
\end{remark}

\begin{remark}{\rm
 As can be conjectured, this method is probably not applicable
to non-generic systems that have some complex attractors besides the
line segment of fixed points (see [Zhang \& Chen, 2005] for
details).}
\end{remark}

Before investigating the structural stability of generic systems, we
first discuss their $\omega $-stability. A dynamical system is
$\omega $-stable if there exists a homeomorphism from its
\textit{non-wandering} set (here it is $I_{o}$) to that of the
system obtained by perturbing it slightly [Smale, 1967].
Hence, a structurally stable dynamical system is necessarily $\omega $%
-stable, but the converse may not be true. Because there always
exists a homeomorphism between two given line segments, it seems
plausible to infer that a generic system is $\omega $-stable.
Unfortunately, it is not true. Observe that Proposition
\ref{paper-prop1}  holds upon the assumption that two given systems
are generic, some other ``unusual'' types of perturbations may lead
to a system that is not generic, thus destroying the  $\omega
$-stability of generic systems. To that end, a new point of view is
required.

Define a family of systems:%
\begin{equation}
\left[
\begin{array}{c}
x_{1}\left( k+1\right) \\
x_{2}\left( k+1\right)%
\end{array}%
\right] =\left\{
\begin{array}{ll}
A_{1}\left[
\begin{array}{c}
x_{1}\left( k\right) \\
x_{2}\left( k\right)%
\end{array}%
\right], & \mbox{if~}\left| x_{1}\left( k\right) -x_{2}\left( k\right)
\right| >\delta , \\
A_{2}\left[
\begin{array}{c}
x_{1}\left( k\right) \\
x_{2}\left( k\right)%
\end{array}%
\right], & \mbox{otherwise,~~~}%
\end{array}%
\right.  \label{family}
\end{equation}%
where%
\begin{equation}
A_{1}=\left[
\begin{array}{cc}
a+b & 0 \\
1 & 0%
\end{array}%
\right] , ~~ A_{2}=\left[
\begin{array}{cc}
a+\lambda b & \left( 1-\lambda \right) b \\
\lambda & \left( 1-\lambda \right)%
\end{array}%
\right] .  \label{system matx.}
\end{equation}
Note that when $\lambda =1$, $A_{2}=A_{1}$, and that this system is a stable
linear system. When $\lambda =0$, $A_{2}=\left[
\begin{array}{cc}
a & b \\
0 & 1%
\end{array}%
\right] $, giving the system defined by Eq. (\ref{modified system}). Hence,
by introducing $\lambda \in \left[ 0,1\right] $, one gets a family of
systems.

It is easy to verify the following result:

\begin{theorem}
\label{ss_thm_2} For each $\lambda \in (0,1]$, system (\ref{family}) has a
unique fixed point $\left( 0,0\right) $.
\end{theorem}

Consider a perturbation of a system in the form of (\ref{modified system})
by choosing a $\lambda$ sufficiently close to but not equal to zero. Theorem %
\ref{ss_thm_2} tells us that the new system has a unique fixed
point. Clearly there are no homeomorphisms between these two systems
since there exist no one-to-one maps from a line segment to a single
point. Moreover, a non-generic system also has a line segment of
fixed points. So, we have the following conclusion:

\begin{theorem}
\label{ss_thm_omega_stability} System (\ref{modified system}) is not $\omega$%
-stable.
\end{theorem}

The following is an immediate consequence.

\begin{corollary}
\label{ss_coro_1} System (\ref{modified system}) is not structurally stable.
\end{corollary}

\begin{remark}{\rm
 The above investigation tells us that the system with $\lambda
=0$ is a rather ill-conditioned one. Will a system with $\lambda \neq 0$ be $%
\omega $-stable (or even structurally stable): We are convinced that
this generally holds, but till now  we have not found a proof.}
\end{remark}

\begin{remark}{\rm
 For a generic system,
no matter it is of type 1 or not, its non-wandering set is just a
line segment of fixed points, therefore its $\omega $-stability is
preserved if a perturbation is on $a$ and $b$, while {\it not}
destroying the structure shown in Eq. (\ref{modified system}).
Consider the discussion in Sec.~2, the system composed of
(\ref{original system})-(\ref{original_constraint}) is proposed for
a new data transmission strategy, hence though the perturbation of
$a$ and $b$ is reasonable, the perturbation of the form
(\ref{family})-(\ref{system matx.}) induced by $\lambda$ does not
make sense physically. Based on this, we can say that $\omega
$-stability is ``robust'' with respect to uncertainty which is
meaningful (The same argument is proposed in Robbin [1972] for
structural stability). However, it is pretty fragile with respect to
such rare uncertainty as that in Eqs. (\ref{family})-(\ref{system
matx.}). In other words, it is robust yet fragile. It is argued in
Doyle [2004] that `robust yet fragile' is the most important
property of complex systems.}
\end{remark}

\begin{remark}
{\rm It follows form the above results that there is a transition
process in the family of systems defined by Eqs.
(\ref{family})-(\ref{system matx.}) as $\lambda $ moves from 1 to 0,
which has been discussed in our another paper [Zhang {\it et al.},
2005].}
\end{remark}

\section{Higher-Order Systems}

In this section, we briefly discuss the high-dimensional cases.

Consider the following two-dimensional system:
\begin{eqnarray*}
x_{1}\left( k+1\right) &=&a_{1}x_{1}\left( k\right) +b_{1}x_{2}\left(
k\right) , \\
x_{2}\left( k+1\right) &=&a_{2}x_{2}\left( k\right) +b_{2}v\left( k\right),
\end{eqnarray*}%
where%
\begin{equation*}
v\left( k\right) =\left\{
\begin{array}{ll}
x_{1}\left( k\right), & \mbox{if} \left| x_{1}\left( k\right) -v\left(
k-1\right) \right| >\delta, \\
v\left( k-1\right), & \mbox{otherwise.}%
\end{array}%
\right.
\end{equation*}%
Introduce a new variable,%
\begin{equation*}
x_{3}\left( k\right) =v\left( k-1\right) ,
\end{equation*}%
and define%
\begin{equation*}
x=\left[
\begin{array}{c}
x_{1} \\
x_{2} \\
x_{3}%
\end{array}%
\right] , ~~ A_{1}=\left[
\begin{array}{ccc}
a_{1} & b_{1} & 0 \\
b_{2} & a_{2} & 0 \\
1 & 0 & 0%
\end{array}%
\right] , ~~ A_{2}=\left[
\begin{array}{ccc}
a_{1} & b_{1} & 0 \\
0 & a_{2} & b_{2} \\
0 & 0 & 1%
\end{array}%
\right] .
\end{equation*}%
Then%
\begin{equation}
x\left( k+1\right) =\left\{
\begin{array}{ll}
A_{1}x\left( k\right), & \mbox{if} \left| x_{1}\left( k\right) -x_{3}\left(
k\right) \right| >\delta, \\
A_{2}x\left( k\right), & \mbox{otherwise.}%
\end{array}%
\right.  \label{2-d system}
\end{equation}

\subsection{Fixed points and switching surfaces}

Suppose that $\left( \bar{x}_{1},\bar{x}_{2},\bar{x}_{3}\right) $ is a fixed
point of system (\ref{2-d system}). Then%
\begin{eqnarray*}
\bar{x}_{1} &=&a_{1}\bar{x}_{1}+b_{1}\bar{x}_{2}, \\
\bar{x}_{2} &=&a_{2}\bar{x}_{2}+b_{2}\bar{x}_{3}, \\
\bar{x}_{3} &=&\bar{x}_{3}.
\end{eqnarray*}%
If $a_{1}\neq 1$ and $a_{2}\neq 1$, then%
\begin{equation}
\bar{x}_{2}=\frac{b_{2}}{1-a_{2}}\bar{x}_{3}, ~~ \bar{x}_{1}=\frac{b_{1}}{%
1-a_{1}}\frac{b_{2}}{1-a_{2}}\bar{x}_{3},  \label{case1a}
\end{equation}%
where%
\begin{equation}
\left| \bar{x}_{3}\right| \leq \frac{\delta }{\left| \frac{b_{1}b_{2}}{%
\left( 1-a_{1}\right) \left( 1-a_{2}\right) }-1\right| }.  \label{case1b}
\end{equation}%
If $a_{1}\neq 1$ and $a_{2}=1$, then%
\begin{equation}
\bar{x}_{3}=0, ~~ \bar{x}_{1}=\frac{b_{1}}{1-a_{1}}\bar{x}_{2}
\label{case2a}
\end{equation}%
and%
\begin{equation}
\left| \bar{x}_{2}\right| \leq \left| \frac{\left( 1-a_{1}\right) \delta }{%
b_{1}}\right| .  \label{case2b}
\end{equation}%
If $a_{1}=1$ and $a_{2}\neq 1$, then%
\begin{equation}
\bar{x}_{2}=0, ~~ \bar{x}_{3}=0  \label{case3a}
\end{equation}%
and%
\begin{equation}
\left| \bar{x}_{1}\right| \leq \delta .  \label{case3b}
\end{equation}%
In all the three cases, fixed points constitute a line segment in $\mathbb{R}%
^{3} $. Note that the case of $a_{1}=1$ and $a_{2}=1$ is contained
in the case defined by Eqs. (\ref{case3a})-(\ref{case3b}).

Next, we consider the first case. Obviously the switching surfaces of this
system are%
\begin{equation*}
x_{1}-x_{3}=\pm \delta ,
\end{equation*}%
hence the two end points of the line of fixed points are
\begin{equation*}
\pm \left( \frac{k_{1}}{k_{1}-1}\delta , \ \frac{b_{2}}{1-a_{2}}\frac{k_{1}}{%
k_{1}-1}\delta , \ \frac{1}{k_{1}-1}\delta \right) ,
\end{equation*}%
where $k_{1}=\frac{b_{1}}{1-a_{1}}\frac{b_{2}}{1-a_{2}}$. They are symmetric
with respect to the origin.

\subsection{An example}
\begin{figure}[tbh]
\epsfxsize=4in
\par
\epsfclipon
\par
\centerline{\epsffile{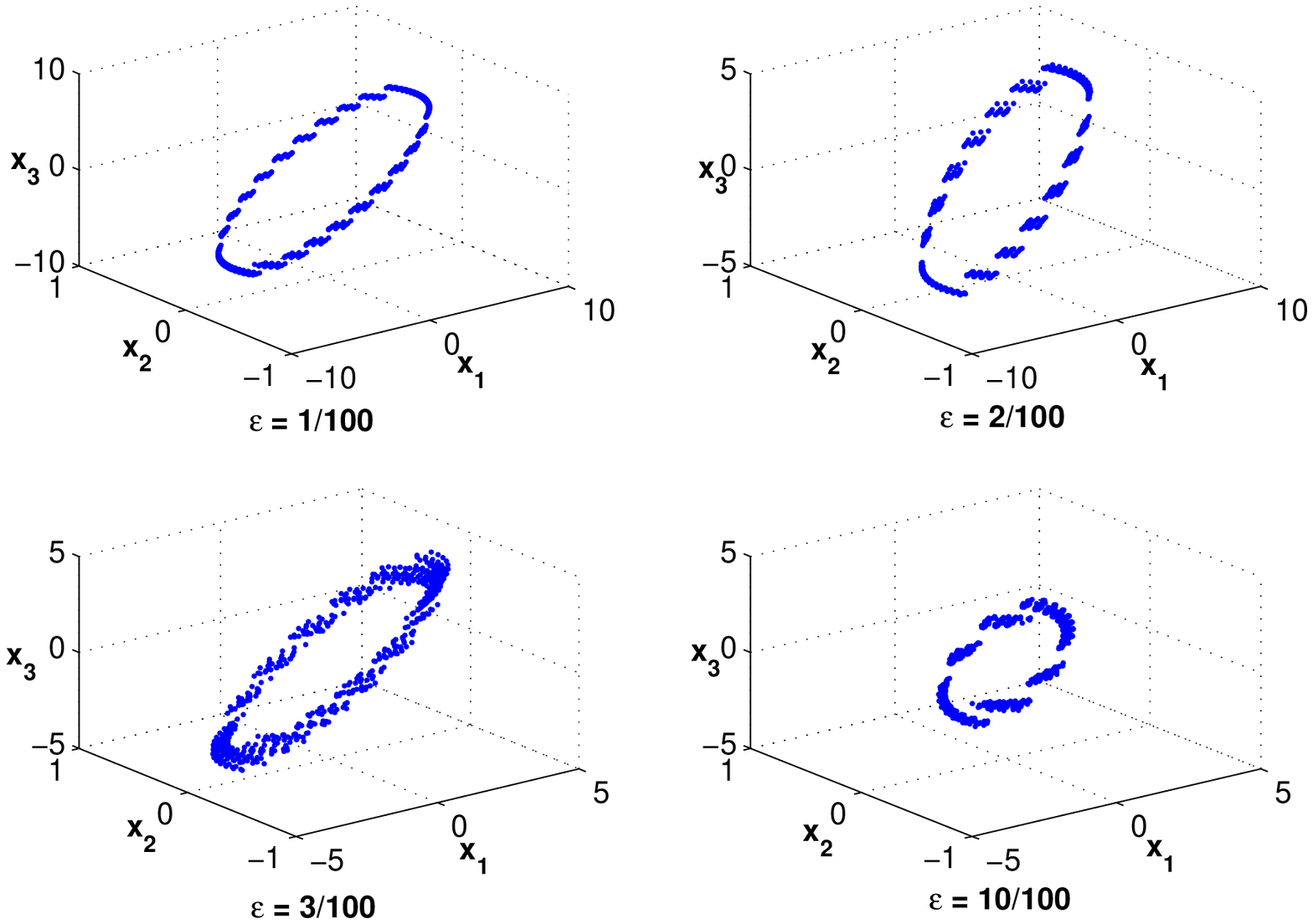}} \caption{Attractors in 3-d:I }
\label{Figure8}
\end{figure}

\begin{figure}[tbh]
\epsfxsize=4.5in
\par
\epsfclipon
\par
\centerline{\epsffile{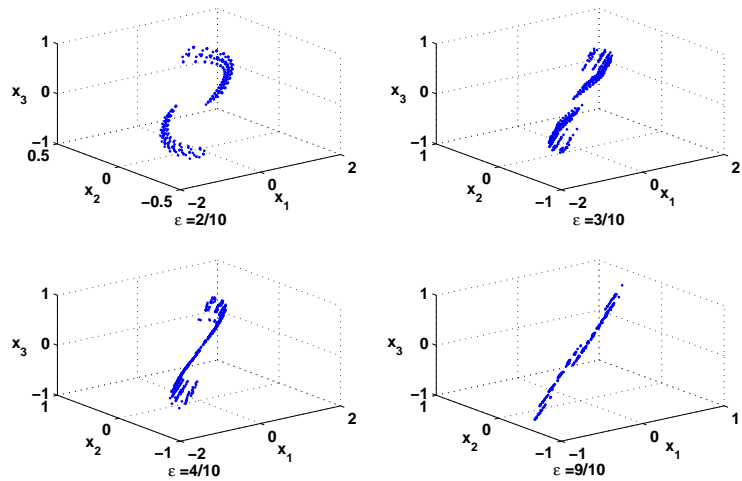}} \caption{Attractors in 3-d:II }
\label{Figure9}
\end{figure}

Based on the analysis in Sec.~4,  one can see that system (\ref{2-d
system}) is not structurally stable. In this section, we illustrate
the complex behavior of this system.
 Consider the following system:
\begin{equation}
x\left( k+1\right) =\left\{
\begin{array}{ll}
A_{1}x\left( k\right), & \mbox{if~}\ \left| x_{1}\left( k\right)
-x_{3}\left( k\right) \right| >1, \\
A_{2}x\left( k\right), & \mbox{otherwise,}%
\end{array}%
\right.  \label{paper2_2-d system}
\end{equation}
where
\begin{equation*}
A_{1}=\left[
\begin{array}{ccc}
1-\epsilon & 1 & 0 \\
-\epsilon/2 & 1 & 0 \\
1 & 0 & 0%
\end{array}%
\right] , ~ A_{2}=\left[
\begin{array}{ccc}
1-\epsilon & 1 & 0 \\
0 & 1 & -\epsilon/2 \\
0 & 0 & 1%
\end{array}%
\right].
\end{equation*}
The variations of the trajectory, starting from $(1, ~ 1/10^5, ~ -1)
$ as $\epsilon$ varies, are plotted in
Figs.~\ref{Figure8}-\ref{Figure9}. One can see the phase transition
process vividly from these figures.

Figs.~\ref{Figure8}-\ref{Figure9} reveal the rich dynamics of a 3-d
system governed by the switching law, which will be our future
research topic.

\section{The Continuous-Time Case}

In this section, we study the continuous-time counterpart of the
discrete-time system (\ref{2-d system}). The first motivation is to
check the data transmission strategy for analog channels; the second
is that discrete-time systems can be regarded as continuous ones if
their base frequencies are much bigger than the network data
transmission rate [Walsh {\it et al.}, 2002]; the third is that a
system governed by this transmission strategy possesses very rich
dynamics, thus it is also interesting in its own right.

\subsection{System setting}

Consider the following system:
\begin{eqnarray}
\dot{x}_{1}(t) &=&a_{1}x_{1}(t)+b_{1}x_{2}(t),  \label{chap8_main_system} \\
\dot{x}_{2}(t) &=&a_{2}x_{2}(t)+b_{2}v(t),  \notag
\end{eqnarray}
where the matrix $A=\left(
\begin{array}{cc}
a_{1} & b_{1} \\
b_{2} & a_{2}%
\end{array}
\right)$ is stable, and the switching law is given by
\begin{equation}  \label{chap8_main_switching}
v(t)=\left\{
\begin{array}{ll}
x_{1}(t), & \mbox{if~} \left| x_{1}(t)-v(t_{-})\right| >\delta , \\
v(t_{-}), & \mbox{otherwise,}%
\end{array}%
\right.
\end{equation}
in which $\delta$ is a positive scalar.

As can be observed, the system governed by Eqs. (\ref{chap8_main_system})-(%
\ref{chap8_main_switching}) is the continuous-time counterpart of system (%
\ref{2-d system}). Note that the system consists of two first-order
ordinary differential equations. Furthermore, if we fix $v(0_{-})$
to be $0$, then the above system is autonomous. Moreover, due to the
switching nature of the system, the vector field of this system may
not be continuous for some set of parameters, not to mention
differentiability. This suggests that the well-known
Poincar$\acute{e}$-Bendixson theorem might not be applicable [Hale
 \& Kocak, 1991], i.e., besides equilibria and periodic orbits,
the $\omega$-limit set of this system may contain other attractors.
This turns out to be true as shown by the following simulations.
However, before doing that, let us first state a
general result regarding the system composed of Eqs. (\ref{chap8_main_system})-%
(\ref{chap8_main_switching}).

\begin{theorem}
\label{chap8_bdd} The trajectories of the system given by Eqs. (\ref%
{chap8_main_system})-(\ref{chap8_main_switching}) are bounded.
Moreover, they converge to the origin as $\delta$ tends to zero.
\end{theorem}

Its proof is omitted due to space limitation.

\subsection{Simulations}

\begin{figure}[t]
\epsfxsize=3.5in
\par
\epsfclipon
\par
\centerline{\epsffile{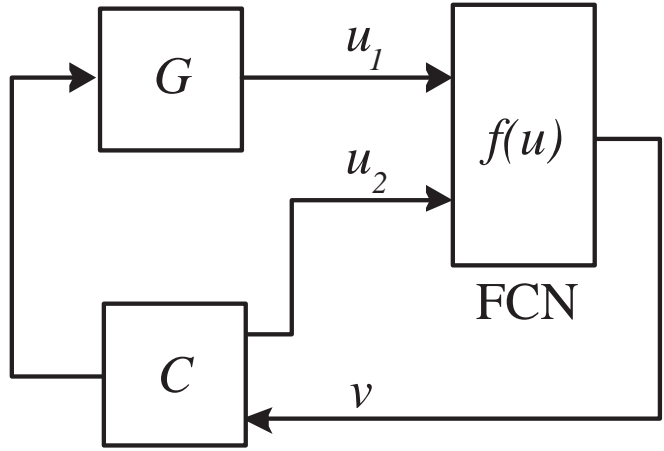}} \caption{A continuous-time
switching system} \label{Figure10}
\end{figure}
Consider the simulink model shown in Fig.~\ref{Figure10}. The system
$G$ and the controller $C$ are modeled by the first and the second
equations in Eq. (\ref{chap8_main_system}), respectively. The function $%
\mathrm{FCN}$, modelling the switching function (\ref{chap8_main_switching}%
), is defined by
\begin{equation}
f(u):=u_1+u_2-u_2\ast (\left| u_1-u_2\right| >\delta)-u_1\ast
(\left| u_1-u_2\right| \leq\delta)~.  \label{Chap8_n}
\end{equation}%
Thus, by letting%
\begin{equation*}
u=\left[
\begin{array}{l}
u_1 \\
u_2%
\end{array}%
\right] ,
\end{equation*}%
one has%
\begin{equation}
v:=f\left( u\right) =\left\{
\begin{array}{ll}
u_1, & \text{if }\left| u_1-u_2\right| >\delta, \\
u_2, & \text{otherwise.}%
\end{array}%
\right.  \label{chap8_constraint}
\end{equation}%
Is the block \textrm{FCN}, the function $f$, well-defined? It
suffices to verify the case at time $0$. Firstly, choose $\left|
x_{1}\left( 0\right) \right| <\delta$. Then $u_1=x_{1}\left(
0\right) $. By simulation, it is found that $v=0$. Secondly, choose
$\left| x_{1}\left( 0\right) \right|
>\delta$. Then $u_1=x_{1}\left( 0\right) $. Simulation shows that $%
v=x_{1}\left( 0\right) $. To simplify the notation, denote $u_2$ at time $0$
by $v\left( 0_{-}\right) $. Summarizing the above, one has%
\begin{eqnarray*}
v\left( 0_{-}\right) &=&0, \\
v\left( 0\right) &=&\left\{
\begin{array}{ll}
x_{1}\left( 0\right) , & \text{if }\left| x_{1}\left( 0\right) -v\left(
0_{-}\right) \right| >\delta, \\
v\left( 0_{-}\right) , & \text{otherwise.}%
\end{array}%
\right.
\end{eqnarray*}%
Similarly,%
\begin{equation*}
v\left( t\right) =\left\{
\begin{array}{ll}
x_{1}\left( t\right) , & \text{if }\left| x_{1}\left( 0\right) -v\left(
0_{-}\right) \right| >\delta, \\
v\left( t_{-}\right) , & \text{otherwise,}%
\end{array}%
~~\mbox{~for~ }t>0,\right.
\end{equation*}%
where $v\left( t_{-}\right) $ is either some previous value of the state $%
x_{1}$, say $x_{1}\left( t-t_{0}\right) $ for some $t_{0}$ satisfying $%
0<t_{0}\leq t$, or $v\left( 0_{-}\right) =0$. Hence, the block
\textrm{FCN} is well-defined.

Next, we find the equilibria of the system. As expected, the
equilibria of the system constitute a line segment, just as in the
discrete-time case. The equilibria are given by
\begin{equation}  \label{Chap8_fixed_points}
\Lambda =\left\{ \left( x_{1}=\frac{b_{1}b_{2}}{a_{1}a_{2}}v, ~ x_{2}=-\frac{%
b_{2}}{a_{2}}v, ~v\right) :\left| v\right| \leq \frac{\delta }{\left| 1-%
\frac{b_{1}b_{2}}{a_{1}a_{2}}\right| }\right\}.
\end{equation}

For the system composed of Eqs. (\ref{chap8_main_system})-(\ref%
{chap8_main_switching}), an interesting question is: Given an initial
condition $x(0)$, will $x(t)$ settle to a certain equilibrium or converge to
a periodic orbit or have more complex behavior? There are two ways of
tracking a trajectory $x(t)$: one is to solve Eqs. (\ref{chap8_main_system}%
)-(\ref{chap8_main_switching}) directly, and the other is by means of
numerical methods. To get an analytic solution, one has to detect the
discontinuous points of the right-hand side of Eq. (\ref{chap8_main_system}%
). We first show that the number of the discontinuous points within any
given time interval is finite.

Start at some time $t_{0}\geq 0$ and assume that $\left( x_{1}\left(
t_{0}\right), x_{2}\left( t_{0}\right) \right) $ and $v\left( t_{0-}\right)
=x_{1}\left( t_{0}\right) $ are given, without loss of generality. Suppose
that the first jump of $v$ is at instant $t_{0}+T$. To be specific in the
following calculation, let $t_{0}=0$. Then
\begin{eqnarray*}
x_{1}\left( T\right) &=&e^{a_{1}T}x_{1}\left( 0\right)
+\int_{0}^{T}e^{a_{1}\left( T-\tau \right) }b_{1}x_{2}\left( \tau \right)
d\tau , \\
x_{2}\left( T\right) &=&e^{a_{2}T}x_{2}\left( 0\right)
+\int_{0}^{T}e^{a_{2}\left( T-\tau \right) }b_{2}x_{1}\left( 0\right) d\tau
\\
&=&e^{a_{2}T}x_{2}\left( 0\right) +\int_{0}^{T}e^{a_{2}u}dub_{2}x_{1}\left(
0\right) .
\end{eqnarray*}%
Consequently,%
\begin{equation}  \label{large}
x_{1}\left( T\right) -x_{1}\left( 0\right) =\left(
e^{a_{1}T}-1\right)x_{1}(0)\left (1-\frac{b_{1}b_{2}}{a_{1}a_{2}}%
\right)+\left( e^{a_{2}T}-e^{a_{1}T}\right) b_{1}\frac{x_{2}\left( 0\right) +%
\frac{b_{2}}{a_{2}}x_{1}\left( 0\right) }{a_{2}-a_{1}}.
\end{equation}
As $T\rightarrow 0$, $e^{a_{1}T}-1\rightarrow 0$, and $
e^{a_{2}T}-e^{a_{1}T}\rightarrow 0$. Moreover, we have already shown
the boundedness of solutions, so there exists a $T^{\ast }>0$ such
that
\begin{equation}
\left| x_{1}\left( T\right) -v\left( 0\right) \right|=\left| x_{1}\left(
T\right) -x_{1}\left( 0\right) \right| <\delta  \label{chap8_terminal}
\end{equation}%
for all $T<T^{\ast }$. Thus, the finiteness of the number of the
discontinuous points within any given time interval is established.

Based on this result, theoretically one can find the analytic solution of
the system. However, it is difficult since the condition in Eq. (\ref%
{chap8_terminal}) has to be checked all the time to determine the
switching time $T$. Moreover, this process depends on the initial
point, $\left( x_{1}\left( t_{0}\right), x_{2}\left( t_{0}\right)
\right) $, which is hard
due to the impossibility of finding the exact $T$ satisfying $%
\left| x_{1}\left( T\right) -v\left( 0\right) \right|=\delta$. This
problem will be addressed in more details in Sec.~6.7.

Another route to study this type of systems is by means of numerical
solutions. In the following, some simulations will be shown to
analyze the complexity of the system depicted in
Fig.~\ref{Figure10}. In all the following trajectory figures, the
horizontal axes stands for $x_1$ and the vertical one is $x_2$.

\subsection{Converging to some fixed point}

\begin{figure}[th]
\epsfxsize=3.5in
\par
\epsfclipon
\par
\centerline{\epsffile{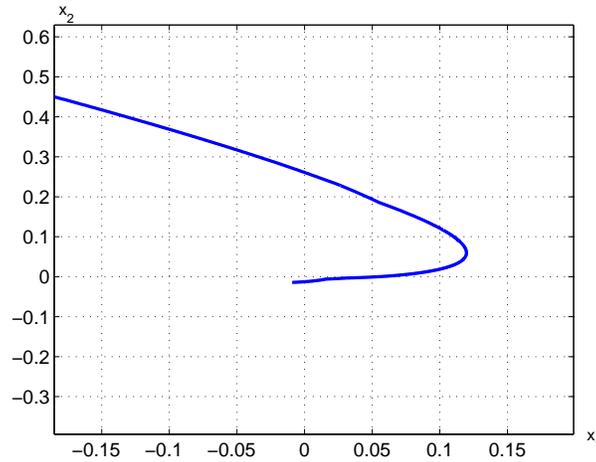}} \caption{Converging to an
equilibrium other than the origin} \label{Figure11}
\end{figure}
Fix those parameters shown in Fig.~\ref{Figure10} to be:
\begin{equation*}
a_{1}=-1,~~ b_{1}=2,~~ a_{2}=-2,~~ b_{2}=-2,
\end{equation*}
and choose an initial condition $(10, -10)$. Then, we get simulation
results shown in Fig.~\ref{Figure11}. One can see that this
trajectory converges to a point specified by Eq.
(\ref{Chap8_fixed_points}), which is close, but not equal, to the
origin.

\subsection{Sensitive dependence on initial conditions}

First, fix system parameters as
\begin{equation}  \label{Chap8_unstable_parmi}
a_{1}=1,~b_{1}=2,~a_{2}=-2,~b_{2}=-2,~\delta=1,
\end{equation}
and note that there is an \textit{unstable} pole in the system $G$.
Suppose
\begin{equation*}
x_{1}\left( 0\right) =2,\quad x_{2}\left( 0\right) =1,
\end{equation*}
and
\begin{equation*}
x_{1}\left( 0\right) =2-10^{-10},\quad x_{2}\left( 0\right) =1.
\end{equation*}
\begin{figure}[th]
\epsfxsize=4in
\par
\epsfclipon
\par
\centerline{\epsffile{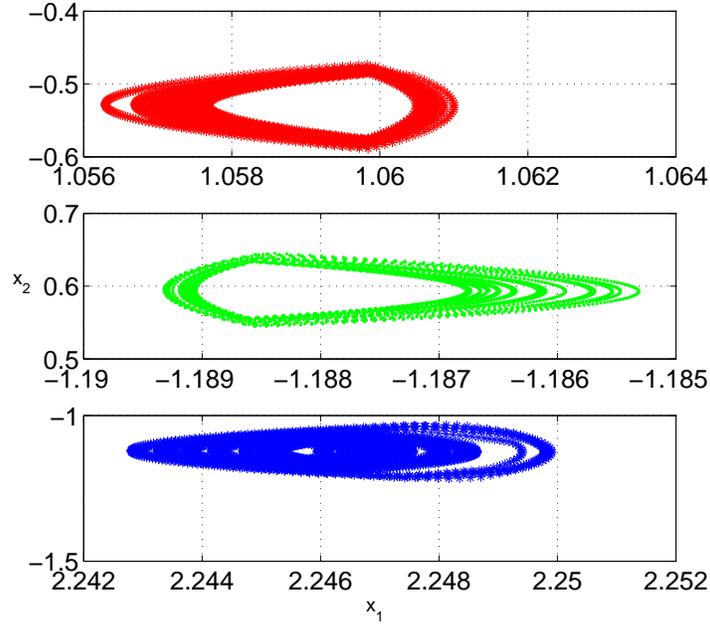}} \caption{Sensitive dependence
on initial conditions (the horizontal axes is $x_1$ and the vertical
one is for $x_2$)} \label{Figure12}
\end{figure}
Then, we get the simulation result shown in Fig.~\ref{Figure12},
where the first two are trajectories from those two sets of initial
conditions given above and the third one is their difference.
Clearly, one can see the sensitive dependence on initial conditions.

\subsection{Coexisting attractors}

\begin{figure}[th]
\epsfxsize=5in
\par
\epsfclipon
\par
\centerline{\epsffile{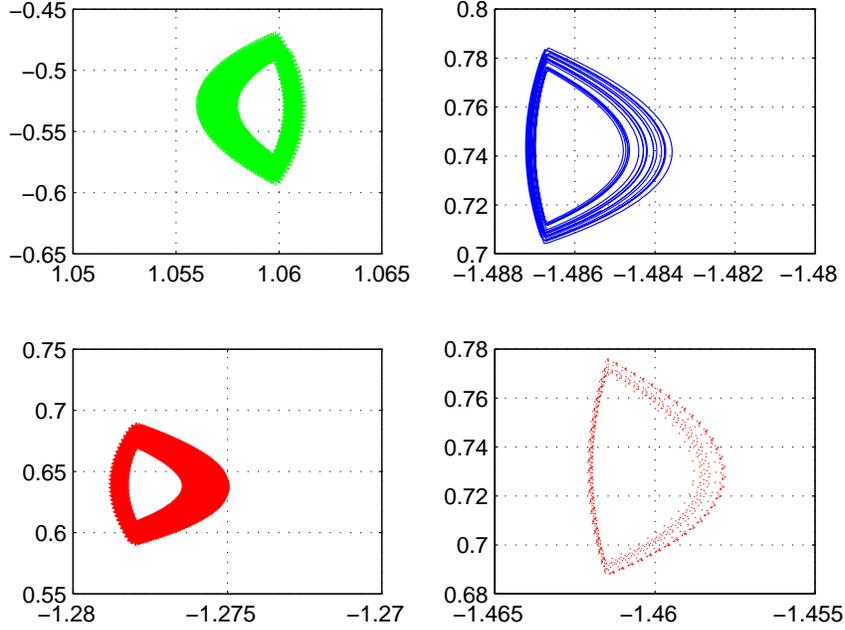}} \caption{Several coexisting
attractors (the horizontal axes is $x_1$ and the vertical stands for
$x_2$)} \label{Figure13}
\end{figure}
Adopt the system parameters as in Eq. (\ref{Chap8_unstable_parmi}).
Then, we have simulation results shown in Fig.~\ref%
{Figure13}, where the largest initial differences of any two such
trajectories is $10^{-3}$. These attractors are all alike; however,
they are located in different positions; that is, they are
coexisting attractors.

\subsection{A periodic orbit}

Now choose
\begin{equation}  \label{Chap8_unstable_parmi2}
a_{1}=-11,~b_{1}=1/4,~a_{2}=10,~b_{2}=1/4-(a_{1}-a_{2})^2,~\delta=1.
\end{equation}
Simulations show that most trajectories behave like the one shown in Fig.~\ref%
{Figure14}, which is periodic.
\begin{figure}[th]
\epsfxsize=6in
\par
\epsfclipon
\par
\centerline{\epsffile{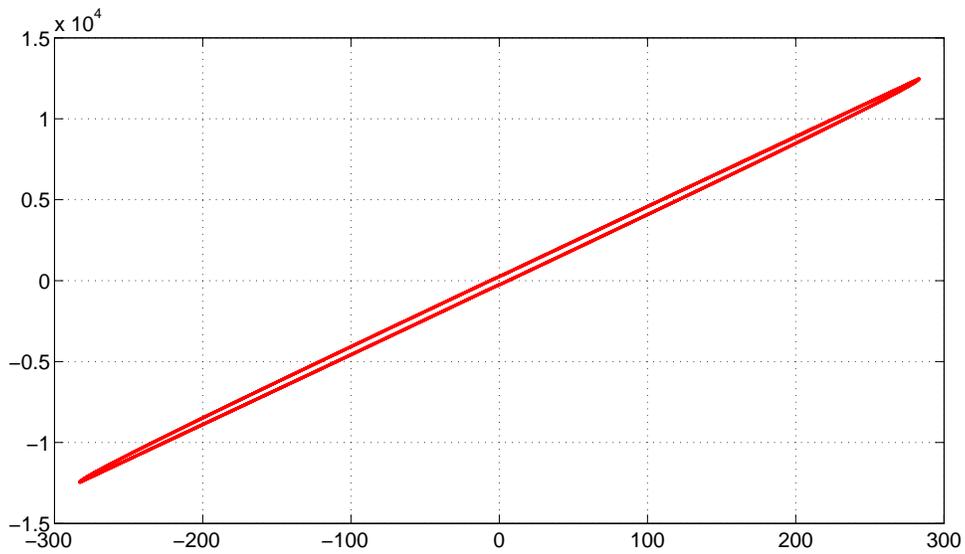}} \caption{A periodic orbit (the
horizontal axes is $x_1$ and the vertical one is for $x_2$)}
\label{Figure14}
\end{figure}

Having observed various complex dynamics possessed by the system
shown in Fig.~\ref{Figure10}, one may ask the following question:

\begin{center}
\textit{Is the complexity exhibited by the system due to numerical errors or
is the system truly chaotic? }
\end{center}

We received the following warning during our simulations using
Simulink: \textit{Block diagram ``A continuous-time switching
system'' contains 1 algebraic loop(s).} This warning is due to the
fact that one of the output of the function block FCN is its own
input. Certainly, this may lead to numerical errors. So, a transport
delay is added to rule out this
possibility. This consideration leads to the following scheme (Fig.~\ref%
{Figure15}):
\begin{figure}[th!]
\epsfxsize=6in
\par
\epsfclipon
\par
\centerline{\epsffile{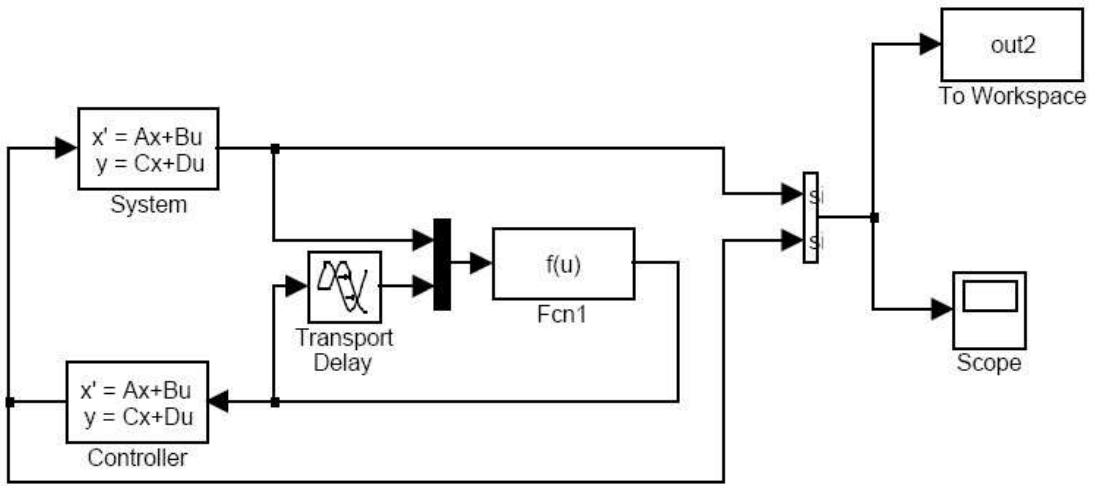}} \caption{A modified
continuous-time switching system} \label{Figure15}
\end{figure}

\begin{figure}[th!]
\epsfxsize=6in
\par
\epsfclipon
\par
\centerline{\epsffile{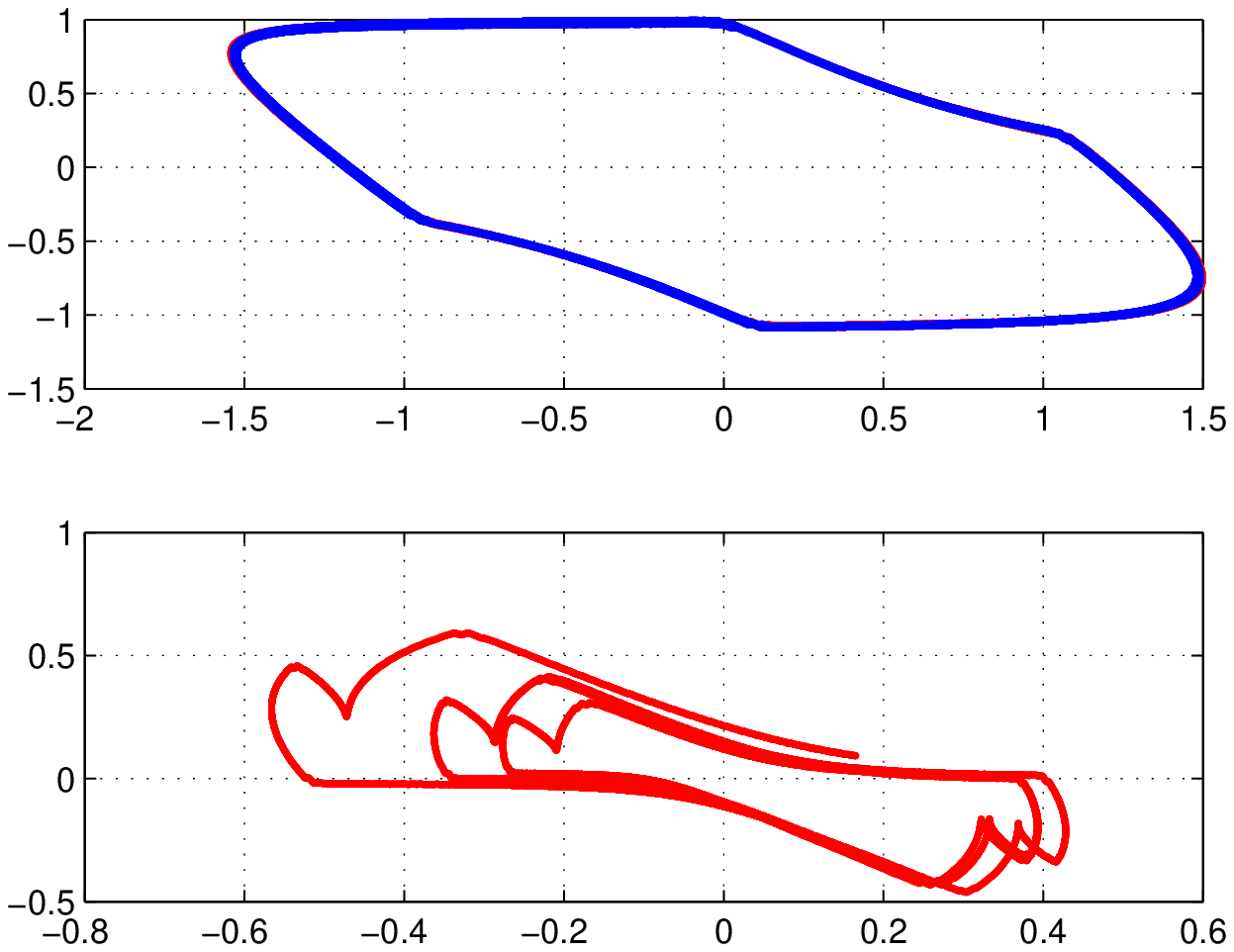}} \caption{Sensitive dependence
on initial conditions (the horizontal axes is $x_1$ and the vertical
one is for $x_2$)} \label{Figure16}
\end{figure}

To correctly implement Eq. (\ref{chap8_main_switching}), the
transport delay T must be small enough. Here, it is fixed to be
$T=5*10^{-2}$. Suppose that system parameters are given by Eq.
(\ref{Chap8_unstable_parmi}) and choose two sets of initial
conditions, $(2,1)$ and $(2-10^{-6},1)$. Then, we get
Fig.~\ref{Figure16}.
According to the upper part of this plot, two trajectories almost coincide;
however, the lower plot clearly reveals sensitive dependence on initial
conditions.

For a sufficiently small transport delay $T$, many simulations show
that the complex attractor is unique, but sensitive dependence on
initial conditions still persists. Apparently, this phenomenon needs
further investigations.

\subsection{Computational Complexity}

We have visualized some complex behaviors of system (\ref%
{chap8_main_system})-(\ref{chap8_main_switching}), but we have not answered
the question posed above. In this section, we study this problem in some
details.

Consider the following system:
\begin{eqnarray}
\dot{x}_{1}\left( t\right) &=&a_{1}x_{1}\left( t\right) +b_{1}x_{2}\left(
t\right) ,  \label{systems} \\
\dot{x}_{2}\left( t\right) &=&a_{2}x_{2}\left( t\right) +b_{2}p,  \notag
\end{eqnarray}%
where%
\begin{equation*}
a_{1}=1,~b_{1}=2,~a_{2}=-2,~b_{2}=-2, ~\delta=1,
\end{equation*}%
and $p$ is a scalar. At $t=0$, let%
\begin{equation*}
x_{1}\left( 0\right) =p,~x_{2}\left( 0\right) =q.
\end{equation*}%
Then, a direct calculation gives
\begin{eqnarray}
x_{1}\left( t\right) &=&2p-\frac{1}{3}e^{t}\left( p-2q\right) -\frac{2}{3}%
e^{-2t}\left( p+q\right) ,  \label{solutions} \\
x_{2}\left( t\right) &=&-p+e^{-2t}\left( p+q\right) .  \notag
\end{eqnarray}%
Suppose $p\neq 2q$. Then, there exists an instant $t_{0}>0$ such that
\begin{equation}
\left| p-x_{1}\left( t_{0}\right) \right| =1.  \label{Chap8_Sec3_switching}
\end{equation}%
Set
\begin{equation*}
p=x_{1}\left( t_{0}\right) .
\end{equation*}%
And then solve equations (\ref{systems}) starting from $\left(
x_{1}\left( t_{0}\right) ,~ x_{2}\left( t_{0}\right) \right) $ at time $%
t_{0} $. Repeat this procedure (update $p$ whenever Eq. (\ref%
{Chap8_Sec3_switching}) is satisfied) to get an analytic solution of the
system starting from $\left( x_{1}\left( 0\right), x_{2}\left(
0\right)\right)=(p,q) $.

Now, it is easy to realize that the complexity may probably be due
to the following reasons:

\begin{itemize}
\item There is an unstable mode in $x_{1}(t)$ in Eq. (\ref{solutions}).

\item It is hard to find the exact switching time, e.g., $t_{0}$ in Eq. (\ref%
{Chap8_Sec3_switching}), even numerically. Because of this,
numerical errors will accumulate and be exaggerated from time to
time by the unstable mode. Further research is required to study the
effect of the accumulated errors on the dynamics of the system.

\item Sec.~6.6 tells us the first item alone can not guarantee complex
behavior.
\end{itemize}

\section{Control Based on the Network Protocol}

Because this research originates from network-based control, in this section
we discuss some control problems under this transmission strategy.

In Zhang \& Chen [2005], concentrated on a scalar case, chaotic
control is investigated. Here, we consider a tracking problem:
suppose the controller $C$ has been designed for the system $G$ in
shown Fig.~\ref{Figure2}, so that the output $y$ tracks the
reference signal $r$. How does the nonlinear constraints $H_{1}$ and
$H_{2}$ affect this tracking problem? We begin with a simple
example.

\begin{example}
{\rm Consider the following discrete-time system $G$:
\[
\frac{0.005z^{-1}+0.005z^{-2}}{1-2z^{-1}+z^{-2}}.
\]%
Note that this system is {\it unstable}. Suppose we have already
designed a controller $K$ of the form
\[
\frac{37.33-33.78z^{-1}}{1-0.1111z^{-1}},
\]%
which achieves step tracking. First, we check the data transmission
strategy by simulating the system shown in Fig.~\ref{Figure3}, with
$r\equiv 1$ but without $H_{2}$ involved therein. Choose $\delta
_{1}=1$. Then, the tracking error is plotted (the $*$ line in
Fig.~\ref{Figure17}).
\begin{figure}[tbh]
\epsfxsize=4in
\par
\epsfclipon
\par
\centerline{\epsffile{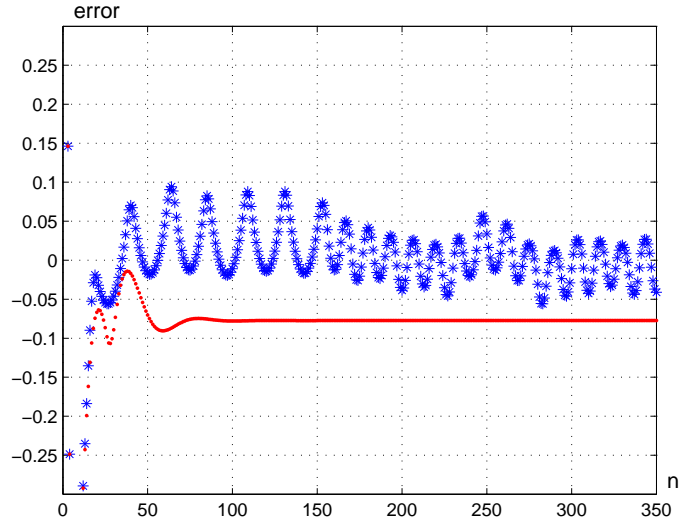}} \caption{Tracking error with
$\delta_{1}=1$} \label{Figure17}
\end{figure}
After that, we
modify the control law as follows:%
\[
v(k)=H_{1}\left( u_{c}\left( k\right) ,v(k-1)\right) =\left\{
\begin{array}{ll}
u_{c}(k), & \mbox{if~}\left| u_{c}\left( k\right) -v\left(
k-1\right)
\right| >1, \\
\varepsilon _{1}v(k-1)+\varepsilon _{2}x\left( k\right) , & \mbox{otherwise,}%
\end{array}%
\right.
\]%
where $\varepsilon _{1}\in \mathbb{R},$  $\varepsilon _{2}\in
\mathbb{R}^{1\times 2}$ are to be determined. Here, we assume that
the state $x$ of the system $G$ is available. When there is no
transmission from $C$ to $G$, instead of simply using the previously
stored control value $v(k-1)$, $\varepsilon _{1}v(k-1)+\varepsilon
_{2}x\left( k\right) $ is used. The reason  is that by adjusting
$\varepsilon _{1}$ and $\varepsilon _{2}$ sensibly we may achieve
better control performance. By selecting
\[
\varepsilon _{1}=0.86, ~ \varepsilon _{2}=\left[ -0.21 ~ 0.21\right]
,
\]%
the tracking error is plotted (the dotted line in Fig.~\ref{Figure17}%
).
\begin{figure}[tbh]
\epsfxsize=4in
\par
\epsfclipon
\par
\centerline{\epsffile{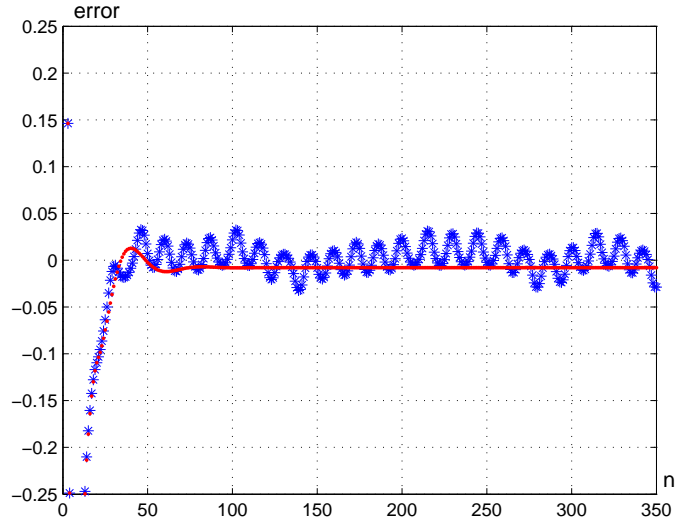}} \caption{Tracking error with
$\delta_{1}=1/2$} \label{Figure18}
\end{figure}
In this simulation, iteration time is 350, so the dropping rate can
also be obtained: the former is $85.14\%$ and the latter is
$96.57\%$. Hence, the modified control law is more effective in
reducing data traffics. The steady-state error shown in
Fig.~\ref{Figure17} under the modified control law is around $0.0774
$. Next, we choose $\delta _{1}=1/2$, and get the result shown in Fig.~\ref%
{Figure18} following the same procedure. The dropping rates are
$83.14\%$ for the original and $96.29\%$ for the modified. In this
case the steady-state tracking error for the modified system is
around $0.0079$. So, by modifying the control law, we increased the
dropping rate therefore reduced the data traffics, and at the same
time improved the performance of the control system.}
\end{example}

In the above example, it is shown that for simple control systems it
is possible to improve the performance of both the network and the
control system by modifying the underlying control law. Clearly, it
is more mathematically involved when one confronts a more complex
system. In the following we transform this problem into an
optimization problem.

Define $\varsigma =\eta -\xi $, and $e_{r}:=e_{c}-e$. By subtracting the
system in (\ref{clsys1}) from that in (\ref{clsys2}), we get
\begin{eqnarray}
\varsigma (k+1) &=&\check{A}\varsigma (k)+\left[
\begin{array}{cc}
B & BD_{d} \\
0 & B_{d}%
\end{array}
\right] \left( \left[
\begin{array}{c}
H_{1}\left( u_{c}\left( k\right) ,v(k-1)\right) \\
H_{2}\left( y_{c}\left( k\right) ,z(k-1)\right)%
\end{array}
\right] -\left[
\begin{array}{c}
u_{c}\left( k\right) \\
y_{c}\left( k\right)%
\end{array}
\right] \right) ,  \notag \\
e_{r}(k) &=&\left[
\begin{array}{cc}
-C & 0%
\end{array}
\right] \varsigma \left( k\right) +z\left( k\right) -y_{c}\left( k\right) .
\label{tracking}
\end{eqnarray}
Then, the tracking error $e_{c}$ of system (\ref{clsys2}) can be obtained
via tracking error $e$ of system (\ref{clsys1}), which is a standard
feedback system.

To study the tracking error $e_{r}$, we employ the $l^{\infty
}$-norm of the system signals [Bamieh, 2003]. According to
(\ref{tracking}), we have
\begin{equation}
\left\| \varsigma \right\| _{\infty }\leq \left\| \left( z^{-1}I-\left[
\begin{array}{cc}
A-BD_{d}C & BC_{d} \\
-B_{d}C & A_{d}%
\end{array}
\right] \right) ^{-1}\left[
\begin{array}{cc}
B & BD_{d} \\
0 & B_{d}%
\end{array}
\right] \right\| _{1}\bar{\delta},  \label{statebound}
\end{equation}
and
\begin{eqnarray}
\left\| e_{r}\right\| _{\infty } &\leq &\left\| \left[
\begin{array}{cc}
-C & 0%
\end{array}
\right] \left( z^{-1}I-\left[
\begin{array}{cc}
A-BD_{d}C & BC_{d} \\
-B_{d}C & A_{d}%
\end{array}
\right] \right) ^{-1}\left[
\begin{array}{cc}
B & BD_{d} \\
0 & B_{d}%
\end{array}
\right] \right\| _{1}  \notag \\
&&~~~\times\left\| \left[
\begin{array}{c}
H_{1}\left( u_{c}\left( k\right) ,v(k-1)\right) \\
H_{2}\left( y_{c}\left( k\right) ,z(k-1)\right)%
\end{array}
\right] -\left[
\begin{array}{c}
u_{c}\left( k\right) \\
y_{c}\left( k\right)%
\end{array}
\right] \right\| _{\infty }+\delta _{2}  \label{trackingbound} \\
&\leq &\bar{\delta}\left\| \left[
\begin{array}{cc}
-C & 0%
\end{array}
\right] \left( z^{-1}I-\left[
\begin{array}{cc}
A-BD_{d}C & BC_{d} \\
-B_{d}C & A_{d}%
\end{array}
\right] \right) ^{-1}\left[
\begin{array}{cc}
B & BD_{d} \\
0 & B_{d}%
\end{array}
\right] \right\| _{1}+\delta _{2}.  \notag
\end{eqnarray}
\begin{figure}[tbh]
\epsfxsize=3.5in
\par
\epsfclipon
\par
\centerline{\epsffile{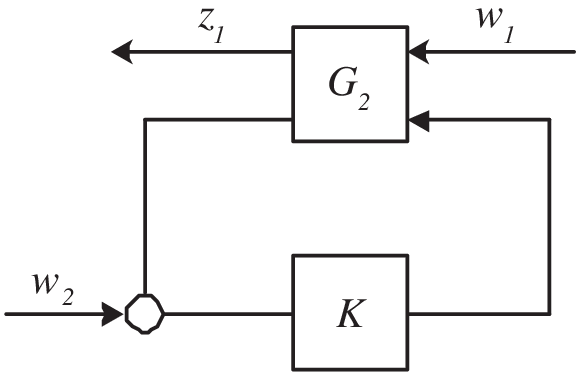}} \caption{An $\ell_{1}$
minimization problem} \label{Figure19}
\end{figure}

\begin{remark}
{\rm Eq. (\ref{trackingbound}) gives an upper bound of the
difference of
the tracking error for the systems shown in Figs. \ref{Figure2}-\ref%
{Figure3}. In light of (\ref{statebound}) and (\ref{trackingbound}),
with $\delta _{1}$ and $\delta _{2}$ fixed, minimizing the size of
an attractor and the tracking error can be converted to the problem
of
designing a controller $K$ that achieves step tracking in Fig.~\ref%
{Figure2} and minimizes $\left\| z_{1}\right\| _{\infty }$ in Fig.~\ref%
{Figure19} simultaneously, with
\begin{equation*}
G_{2}=\left[
\begin{array}{c|cc}
A & B & B \\ \hline
-C & 0 & 0 \\
-C & 0 & 0%
\end{array}
\right] , \ K=\left[
\begin{array}{c|c}
A_{d} & B_{d} \\ \hline
C_{d} & D_{d}%
\end{array}
\right] , \ \left\| w_{1}\right\| _{\infty }\leq 1, \ \left\| w_{2}\right\|
_{\infty }\leq 1.
\end{equation*}
For this multiple-objective control problem, LMI techniques can be
applied. More specifically, by parameterizing all stabilizing
controllers, the step tracking problem has
an equality constraint for the $l^{1}$control problem shown in Fig.~\ref%
{Figure19}, which can be modified as an LMI minimization problem
(see [Chen \& Francis, 1995] and [Chen \& Wen, 1995] for more
details).}
\end{remark}

\section{Conclusions}

In this paper, we have generalized the results of Zhang \& Chen
[2005] in the following ways: 1) We have constructed first-return
maps of the nonlinear systems in Zhang \& Chen [2005] and derived
existence conditions for periodic orbits and studied their
properties. 3) We have formulated the involving systems as hybrid
systems, and proved that this type of hybrid systems is not
structurally stable. 4) We have examined higher-dimensional models
with detailed studies of the existence of periodic orbits. 5) We
have investigated a class of continuous-time hybrid systems as the
counterparts of the discrete-time systems. 6) We have proposed new
controller design methods based on this network transmission
strategy for improving control performance of individual systems as
well as the whole network.

Interestingly, one application of this network data transmission
strategy is the so-called limited communication control in control
and coordination of multiple subsystems. One example is: A single
decision maker controls many subsystems over a communication channel
of a finite capacity, where the decision maker can control only one
subsystem at a time. Let us consider the following situation:
Suppose there are several systems sharing a common communication
channel, where at each transmission time only one system can send a
signal. Is it possible that each subsystem adopts the transmission
strategy proposed here so that the whole system can achieve some
desired system performance? Note that under the proposed
transmission strategy, each system just sends ``necessary'' signals,
leaving communication resources to the others to use. So, if we
design the \textit{transmission sequence} carefully, the whole
system might perform well. A similar but essentially different
problem was discussed in Hristu \& Morgansen [1999], which is an
extension of the work of Brockett [1995]. The problem studied
therein is: Given a set of control systems controlled by a single
decision maker, which can communicate with only one system at a
time, design a communication sequence so that the whole network is
asymptotically stable. Using augmentation, this problem can be
converted to a mathematical programming problem for which some
algorithms are currently available. Here, under the proposed
transmission strategy, the communication strategy depends severely
on the control systems. Hence the communication sequence depends
explicitly on all subsystems, adding more constraints to the design
of the communication sequence. This important yet challenging
problem will be our future research topic.

\section{Acknowledgement}

This work was partially supported by NSERC. G. Zhang is grateful to
the discussions with Dr. Michael Y. Li and Dr. Y. Lin. M.B. D'Amico
appreciates the financial support of SGCyT at the Universidad
Nacional del Sur, CONICET, ANPCyT (PICT -11- 12524) and the City
University of Hong Kong (CERG CityU 1114/05E).

\noindent {\bf References}

Bamieh, B. [2003] ``Intersample and finite wordlength effects in
sampled-data problems,'' {\it IEEE Trans. Automat. Contr.} {\bf
48}(4), 639-643.

Brockett, R. \& Liberzon, D. [2000] ``Quantized feedback
stabilization of linear systems,'' {\it IEEE Trans. Automat. Contr.}
 {\bf 45}(7), 1279-1289.

Brockett, R. [1995] ``Stabilization of motor networks,'' in {\it
Proc. of IEEE Conf. Decision and Control}, pp. 1484-1488.

Cervin, A., Henriksson, D., Lincoln, B., Eker, J., \& Arz$%
\acute{e}$n, K. [2003] ``How does control timing affect
performance?'' {\it IEEE Control Systems Magazine} {\bf 23}(1),
16-30.

Chen, T. \& Francis, B. [1995] {\it Optimal Sampled-Data  Control
Systems}(Springer:London).

Chen, X. \& Wen, J. [1995] ``A linear matrix inequality approach to
the discrete-time mixed $l_{1}/\mathcal{H}_{\infty }$ control
problem,'' in {\it Proc. of IEEE Conf. Decision and Control}, pp.
3670-3675.

Delchamps, D. [1988] ``The stabilization of linear systems with
quantized feedback,'' in {\it Proc. of IEEE Conf. Decision and
Control}, pp. 405-410.

Delchamps, D. [1989] ``Controlling the flow of information in
feedback systems with measurement quantization,'' in {\it Proc. IEEE
Conf. Decision and Control}, pp. 2355-2360.

Delchamps, D. [1990] ``Stabilizing a linear system with quantized
state feedback,'' {\it IEEE Trans. Automat. Contr.} {\bf 35}(8),
916-924.

Doyle, J. [2004] ``Complexity,'' Presented at Georgia Institute of
Techology.

Elia, N. [2004] ``When Bode meets Shannon: control-oriented feedback
communication schemes,'' {\it IEEE Trans. Automat. Contr.} {\bf
49}(9), 1477-1488.

Fagnani, F. \& Zampieri, S. [2003] ``Stability analysis and
synthesis for scalar linear systems with a quantized feedback,''
{\it IEEE Trans. Automat. Contr.} {\bf 48}(8), 1569-1583.

Fagnani, F. \& Zampieri, S. [2004] ``Quantized stabilization of
linear systems: complexity versus performance,'' {\it IEEE Trans.
Automat. Contr.} {\bf 49}(9), 1534-1548.

Goodwin, G., Haimovich, H., Quevedo, D. \& Welsh, J. [2004] ``A
moving horizon approach to networked control system design,'' {\it
IEEE Trans. Automat. Contr.} {\bf 49}(9), 1427-1445.

Hale, J. \& Kocak, H. [1991] {\it Dynamics and Bifurcations}
(Springer-Verlag).

Hristu, D. \& Morgansen, K. [1999] ``Limited communication
control,'' {\it Systems \& Control Letters} {\bf 37}(4), 193-205.

Ishii, H. \& Francis, B. [2002] ``Stabilization with control
networks,'' {\it Automatica} {\bf 38}(10), 1745-1751.

Khalil, H. [1996] {\it Nonlinear Systems}(Prentice Hall).

Krtolica, R.,  \"{O}zg\"{u}ner, \"{U}., Chan, D., G\"{o}%
ktas, G., Winkelman, J. \& Liubakka, M. [1994] ``Stability of linear
feedback systems with random communication delays,'' {\it Int. J.
Control} {\bf 59}, 925-953.

Lasota, A. \& Yorke, J. [1973] ``On the existence of invariant
measures for piecewise monotonic transformations,'' {\it Trans.
Amer. Math. Soc.} (186), 481-488.

Li, T. \& Yorke, J. [1978] ``Ergodic transformations from an
interval to itself,'' {\it Trans. Amer. Math. Soc.} (235), 183-192.

Montestruque, L. \& Antsaklis, P. [2004] ``Stability of model-based
networked control systems with time-varying transmission times,''
{\it IEEE Trans. Automat. Contr.} {\bf 49}(9), 1562-1572.

Murray, R., ${\AA}$str$\ddot{o}$m, K.,  Boyd, S., Brockett, R. \&
Stein, G. [2003] ``Future directions in control in an
information-rich world,'' {\it IEEE Control Systems Magazine} {\bf
23}(2), 20-33.

Nilsson, J., Bernhardssont, B.  \& Witenmark, B. [1998] ``Stochastic
analysis and control of real-time systems with random time delays,''
{\it Automatica} {\bf 34}(1), 57-64.

Nesic, D \&  Teel, A. R. [2004] ``Input output stability properties
of networked control systems,'', {\it IEEE Trans. Automat. Contr.}
{\bf 49}(10), 1650-1667.

Octanez, P., Monyne, J., \& Tilbury, D. [2002] ``Using deadbands to
reduce communication in networked control systems,'' in {\it Proc.
of American Control Conference}, pp. 3015-2020.

Raji, R. [1994] ``Smart networks for control,''{\it IEEE Spectrum},
{\bf 31}, 49-53

Robbin, J. W. [1972] ``Topological conjugacy and structural
stability for discrete dynamical systems,'' {\it Bell. Amer.Math.
Soc.} (78), 923-952.

Smale, S. [1967] ``Differentible dynamical systems,'' {\it Bull.
Amer. Math. Soc.}  (73), 747-817.

Tatikonda, S. \& Mitter, S. [2004] ``Control under communication
constraints,'' {\it IEEE Trans. Auto. Contr.}  {\bf 49}(9),
1056-1068.

Tatikonda, S. \& Mitter, S. [2004b] ``Control over noisy channels,''
{\it IEEE Trans. Auto. Contr.} {\bf 49}(9), 1196-1201.

Wong, W. \& Brockett, R. [1999] ``Systems with finite communication
bandwidth constraints, part II: stabilization with limited
information feedback,'' {\it IEEE Trans. Auto. Contr.} {\bf 44}(5),
 1049-1053.

Walsh, G., Beldiman, O., \& Bushnell, L. [1999] ``Error encoding
algorithms for networked control systems,'' in Proc. of IEEE Conf.
Decision and Control, pp. 4933-4938, 1999.

Walsh, G., Beldiman, O., \& Bushnell, L. [2001] ``Asymptotic
behavior of nonlinear networked control systems,'' {\it IEEE, Trans.
Automat. Contr.} {\bf 46}(6), 1093 -1097.

Walsh, G., Beldiman, O., \& Bushnell, L. [2002] ``Stability analysis
of networked control systems,'' {\it IEEE Trans. Contr. Syst.
Techno.} {\bf 10}(3), 438-446.

Walsh, G., Hong, Y., \& Bushnell, L. [2002b] ``Error encoding
algorithms for networked control systems,'' {\it Automatica} {\bf
38}(2), 261-267.

Walsh, G. \& Ye, H. [2001] ``Scheduling of networked control
systems,'' {\it IEEE Control Systems Magazine} {\bf 21}(1), 57-65.

Yue, D., Han, Q., \& Lam, J. [2005] ``Network-based robust $%
H_{\infty}$ control of systems with uncertainty,'' {\it Automatica}
{\bf 41}(6), 999-1007.

Zhang, G. \& Chen, T. [2005] ``Networked control systems: a
perspective from chaos,'' {\it Int. J. Bifurcation and Chaos}, {\bf
15}(10), 3075-3101.

Zhang, G., Chen, G., Chen, T. \& Lin, Y. [2005] ``Analysis of a type
of nonsmooth dynamical systems,'' {\it Chaos, Solitons \& Fractals}
{\bf 30}, 1153-1164.

\end{document}